\documentclass[twocolumn,amsmath,amssymb,aps,prc,nofootinbib]{revtex4-2}
\usepackage{graphicx,here}
\usepackage{dcolumn,bm,hyperref,footnote}
\usepackage{float}
\usepackage[dvipsnames,svgnames,x11names]{xcolor}
\usepackage[authormarkup=none,defaultcolor=BrickRed]{changes}
\usepackage{multirow}
\usepackage{lipsum}
\usepackage{tikz}
\usepackage{booktabs}
\usepackage{afterpage}
\begin{document}
\newcommand {\nc} {\newcommand}
\nc {\cgmf} {$\texttt{CGMF}$}

\title{Reducing parametric uncertainties through information geometry methods}

\author{M. Imbri{\v s}ak}
\email{marko.imbrisak@gmail.com}
\affiliation{Los Alamos National Laboratory, Los Alamos, NM, 87545, USA}
\author{A. E. Lovell}
\affiliation{Los Alamos National Laboratory, Los Alamos, NM, 87545, USA}
\author{M. R. Mumpower}
\affiliation{Los Alamos National Laboratory, Los Alamos, NM, 87545, USA}

\date{\today}

\begin{abstract}
Information geometry is a study of applying differential geometry methods to challenging statistical problems, such as uncertainty quantification. In this work, we use information geometry to study how measurement uncertainties in pre-neutron emission mass distributions affect the parameter estimation in the Hauser-Feshbach fission fragment decay code, \cgmf{}. We quantify the impact of reduced uncertainties on the pre-neutron mass yield of specific masses to these parameters, for spontaneous fission of ${}^{252}$Cf, first using a toy model assuming Poissonian uncertainties, then an experimental measurement taken from G\"o\"ok et al., 2014 in EXFOR.
We achieved a reduction of up to $\sim15\%$ in \cgmf{} parameter errors, predominantly in $w_0^{(1)}$ and $w_1^{(0)}$.
\end{abstract}

\maketitle

\section{Introduction\label{sec:intro}}

The fission of a heavy nucleus is a process in which the parent nucleus undergoes a complex process of splitting into two or more lighter fission fragments. Immediately following the scission of the compound nucleus, prompt neutrons and $\gamma$ rays are emitted on a very short timescale, reaching either a ground state or long-lived isomeric state. On longer timescales, these nuclei can further $\beta$ decay, emitting delayed neutrons and $\gamma$ rays. The emissions of these prompt and delayed neutrons and $\gamma$ rays depend on the initial conditions of the fission fragments, e.g. their mass, charge, excitation energy, spin, and parity. Because of the short timescales of the scission process, these initial conditions--before prompt emission--are impossible to determine directly from experiment and must be inferred through subsequent prompt and delayed measurements or from theory.

 In nuclear fission reactions, the parent nucleus undergoes a complex process of splitting into two or more lighter nuclei. This process results in a wide distribution of fission fragments characterized by variations in mass, charge, and kinetic energy. The pre-scission phase, which occurs before the actual splitting, has been the focus of extensive theoretical investigations.
 Different models can be used to describe this phase. For example, in the macroscopic-microscopic approach \cite{Randrup:2011,Moller:2015,Mumpower:2020}, the nucleus is treated as a quantum charged liquid drop while the purely microscopic approach  \cite{Schunck:2014,Bulgac:2016} starts from phenomenological descriptions of nucleon-nucleon forces. 
Recently, \cite{Regnier:2016,Sierk:2017,Usang:2017,Regnier:2019}, researchers have made predictions regarding primary fission fragment yields. These predictions take into account the pre-neutron fission fragment mass distribution. 

 The \cgmf{} code \cite{Talou2021} models the de-excitation of fission fragments on an event-by-event basis using the Hauser-Feshbach statistical model of nuclear reactions \cite{Hauser1952}. It samples initial conditions of the fission fragments in mass, charge, total kinetic energy, spin, and parity, that have been parametrized and fit to available experimental data. Prompt neutrons and $\gamma$ rays are emitted probabilistically, and their energies and momenta are recorded for each event, along with the initial conditions of the fission fragments. These event histories can be used to analyze various mean values, distributions, and correlations of neutrons, $\gamma$ rays, and fission fragments. Currently, \cgmf{} models binary fission while accounting for multi-chance fission and the emission of pre-fission neutrons.
 
In parallel with the development of \cgmf{} and other similar codes \cite{FREYA1, FREYA2,FIFRELIN,Okumura2018,BeoH}, it is necessary to conduct modern fission experiments (see e.g. \cite{Geppert2019,Gook2014,Hambsch1997,Hensle2020}) to better understand the fission process as well as constrain phenomenological inputs to theoretical models. The impact of the experimental uncertainties on these inputs should be quantified, but to date, there have been very few studies of this nature, e.g. \cite{Jaffke2018,Randrup2019}. However, uncertainty quantification as a whole has been increasing across the nuclear theory community, e.g. \cite{Schindler2009,Dobaczewski2014,Perez2015,Schunk2015,McDonnell2015,Furnstahl2015,Furnstahl2015a,Wesolowski2016,Melendez2017,Lovell2018,CatacoraRios2019,Phillips2021,Semposki2022,Kejzlar2023,Sharma2024}.

Information geometry is an interdisciplinary field that introduces methods of differential geometry to statistical problems like uncertainty quantification \cite{Amari1982,Amari2016}. While its initial applications focused on machine learning, neural networks \cite{Ma1997,Amari1998}, and to thermodynamics problems \cite{Ruppeiner1979}, its methods have started being successfully applied to various complex problems in biology \cite{Transtrum2015}, chemistry \cite{Transtrum2016} and physics \cite{Niksic2016,Tisanic2020}. 
In the context of uncertainty quantification in nuclear physics, information geometry has been used in the framework of nuclear energy density functionals to reduce model parameters and analyze self-consistent models \cite{Niksic2016,Niksic2017,Imbrisak2023a,Imbrisak2023b}, offering a bridge between microscopic details and statistical insights.
Motivated by this successful application in energy density functionals, we investigate its applications to uncertainty quantification of \cgmf{} model parameters.

 In Sec. \ref{sec:update}, we present our method for updating measurement errors based on theoretical uncertainties constrained using information geometry. Next, in Sec. \ref{sec:cgmf_description}, we give a brief overview of the method that \cgmf{} uses to sample fission fragment mass distributions. Then, we describe the use of these \cgmf{}-sampled distributions to test our method on well-behaved Poissonian errors in Sec. \ref{sec:poisson} and demonstrate its application to real data in Sec.~\ref{sec:realres}. Finally, we conclude in Sec. \ref{sec:conclusion}.
\section{Model-based measurement error updates \label{sec:update}}
 We use the following conventions to make the procedure more transparent: (1) the measurements are indexed by letters $i,j,\cdots$, (2) the model parameters are indexed by Greek letters, (3) the eigenvalues of the parameter covariance matrix are indexed by letters $a,b,\cdots$, (4) the space of residuals is labeled $\mathcal{D}$, the space of the model parameters $\mathcal{M}$ and the space of covariance matrix eigenvalues by $\Lambda$, and (5) the Einstein summation convention is used whenever possible. 
{\color{black}In the standard minimization procedure, we model a set of $N$ measurements $\{y^i\}_{i\in\{1,\cdots,N\}}=\{y^1,\cdots, y^N\}$ with uncertainties $\{\sigma^i\}_{i\in\{1,\cdots,N\}}=\{\sigma^1,\cdots, \sigma^N\}$ by a model $f^1(\theta),\cdots, f^N(\theta)$ described by parameters $\{\theta^\mu\}_{\mu\in\{1,\cdots,M\}}=\{\theta^1,\cdots,\theta^M\}$.} We assume that the standardized differences between measurements and model evaluations, i.e. residuals, follow the normal distribution and are identically distributed
\begin{align}
     r^i(\theta)&=\frac{y^i-f^i(\theta)}{\color{black}{\sigma^i}}\sim\mathcal{N}(0,1)\color{black}{.}
\end{align}
The best-fitting parameters $\theta$ are found by minimizing the $\chi^2$ value 
\begin{align}
    \chi^2(\theta)&=\frac{1}{2}\sum\limits_{i=1}^N \left(r^i(\theta)\right)^2,
\end{align}
which is a result of the standard likelihood maximization technique. This technique also yields a theoretical lower bound to the parameter errors. {\color{black} By  the Cram\'{e}r-Rao bound \citep[see, e.g.,][]{Cramer99}}, the inverse of the parameter covariance matrix is bounded by the Fisher information matrix \cite{Nielsen2013}
\begin{align}
    g_{\mu\nu}&=\color{black}{\sum\limits_{i=1}^N \partial_\mu\frac{f^i}{\sigma^i}\partial_\nu\frac{f^i}{\sigma^i},}
\end{align}
{\color{black} where we use the shorthand notation $\partial_\mu=\frac{\partial}{\partial\theta^{\mu}}$.}
In this section, we consider the inverse problem: if we would like to decrease the uncertainties on the model parameters, for which measurements do we need to lower measurement uncertainties. We  start with a brief overview of the differential geometry procedures in section \ref{sec:difgeo} and then use them to propose the measurement uncertainty update estimate in section \ref{sec:difgeo2}.

\subsection{Differential geometry methods\label{sec:difgeo}}
 
The eigenvalues of the rank-2 (covariant-index) metric tensor {\color{black}$g_{\mu\nu}$} are given by $\lambda_a$. {\color{black}The metric tensor in the space of eigenvalues, $\Lambda$, is just the diagonal matrix, $g^{(\Lambda)}_{ab}$. 
Its inverse is a rank-2 contravariant tensor, $g^{(\Lambda)}{}^{ab}$.} 
     In the particular case in which we want to describe the eigenvalues of the covariance matrix, $\lambda_a=1/\sigma^2_a$
     \begin{align}
    g^{(\Lambda)}{}^{ab}&=\delta^{ab}\frac{1}{\lambda_a}=\sigma_a^2\delta^{ab}.
    \end{align}
    We now transfer the metric tensor to the model space $\mathcal{M}$ by means of the matrix of eigenvectors, $V_\mu{}^a$, as computed when solving the eigenvalue problem. This transformation yields the metric for the model space $\mathcal{M}$:
    \begin{align}
    g^{(\mathcal{M})}_{\mu\nu}&=[V g^{(\Lambda)
} V^T]_{\mu\nu}\\
    &={\color{black}\sum\limits_{a=1}^M\sum\limits_{b=1}^M}V_\mu{}^{a}\lambda_a \delta_{ab} [V^T]^b{}_\nu\\
    &={\color{black}\sum\limits_{a=1}^M}V_\mu{}^{a}V_\nu{}^{a}\lambda_a 
    \end{align}
    The indices of $g$ and $V$ can be raised by the metric tensor inverse, $V^{\mu a}=g^{(\mathcal{M})}{}^{\mu\nu}V_{\nu}{}^{a}$, keeping in mind that the matrix $g^{\mu\nu}$ needs to correspond to the inverse of $g$. In the language of tensor algebra, this results in the following transformations, as expected from linear algebra
    \begin{align}
    g^{(\mathcal{M})}{}^{\mu\nu}&=g^{(\mathcal{M})}{}^{\mu\rho}g^{(\mathcal{M})}{}^{\nu\sigma}g^{(\mathcal{M})}_{\rho\sigma}\\
    &=[V g^{(\Lambda)}{}^{-1}V^T]^{\mu\nu}\\
    &={\color{black}\sum\limits_{a=1}^M\sum\limits_{b=1}^M}V^\mu{}_{a}V^\nu{}_{b}\frac{\delta^{ab}}{\lambda_a}
    \end{align}
    In the language of differential geometry, covariant rank-1 tensors are represented by vectors in the cotangent bundle (e.g., $T^*\mathcal{M}$) and contravariant rank-1 tensors are represented by vectors in the tangent bundle (e.g., $T\mathcal{M}$). The inverse metric tensor is an element of a direct product of two tangent bundles, \begin{align}g^{-1}{}^{(\Lambda)}={\color{black}\sum\limits_{a=1}^M} \frac{1}{\lambda_a}\frac{\partial}{\partial V^a}\otimes \frac{\partial}{\partial V^a}\in T\Lambda\times T\Lambda.\end{align} The metric tensor is represented by a functional acting on elements of $T\Lambda\times T\Lambda$, i.e., \begin{align}g^{(\Lambda)}={\color{black}\sum\limits_{a=1}^M} \lambda_a dV^a\otimes dV^a\in T^*\Lambda\times T^*\Lambda. \end{align} 

    {\color{black}The raising and lowering of indices are represented by the musical isomorphisms. The raising of indices is performed by the sharp operator $\#: T^*\Lambda\to T\Lambda$, while the lowering of indices is performed by the flat operator, $\flat: T\Lambda\to T^*\Lambda$. In appendix \ref{sec:musiciso} we give a brief description of these operators.} 
    
    \begin{figure}[H]
    \centering\resizebox{\columnwidth}{!}{
    \begin{tikzpicture}
    \node (a) at (0,3)   {$T^*\mathcal{D}$}; 
    \node (b) at (0,0)   {$T\mathcal{D}$}; 
    \node (c) at (3,3)   {$T^*\mathcal{M}$}; 
    \node (d) at (3,0)   {$T\mathcal{M}$}; 
    \node (e) at (6,3)   {$T^*\Lambda$}; 
    \node (f) at (6,0)   {$T\Lambda$}; 
    \draw [->] ([xshift=1mm] a.south) -- node[right=1mm] {$\sharp$} ([xshift=1mm] b.north);
 \draw [->,color=red] ([xshift=-1mm] b.north) -- node[left=1mm] {$\flat$} ([xshift=-1mm] a.south);

\draw [->] ([xshift=1mm] c.south) -- node[right=1mm] {$\sharp$} ([xshift=1mm] d.north);
 \draw [->] ([xshift=-1mm] d.north) -- node[left=1mm] {$\flat$} ([xshift=-1mm] c.south);

\draw [->] ([xshift=1mm] e.south) -- node[right=1mm] {$\sharp$} ([xshift=1mm] f.north);
 \draw [->] ([xshift=-1mm] f.north) -- node[left=1mm] {$\flat$} ([xshift=-1mm] e.south);

 \draw [->] (a) -- node[above=1mm] {$r^*$} (c);
 \draw [->,color=red] (d) -- node[above=1mm] {$r_*$} (b);

 \draw [->,color=red] ([yshift=1mm]f.west) -- node[above=1mm] {$V_*$} ([yshift=1mm]d.east);
 \draw [<-] ([yshift=-1mm]f.west) -- node[below=1mm] {$V^T_*$} ([yshift=-1mm]d.east);

 \draw [->] ([yshift=1mm]e.west) -- node[below=2mm] {$V^{*}$} ([yshift=1mm]c.east);
 \draw [<-] ([yshift=-1mm]e.west) -- node[above=2mm] {$V^{T*}$} ([yshift=-1mm]c.east);
\end{tikzpicture}  }
    \caption{Schematic representation of mappings used in section \ref{sec:update}. The mapping $\flat r_* V_*$ used in section \ref{sec:difgeo2} is shown in red.}
    \label{fig:scheme}
\end{figure}
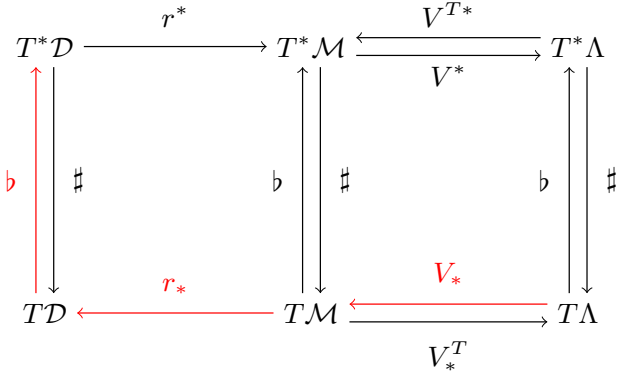

\subsection{Which measurements to conduct to reduce parameter errors\label{sec:difgeo2}}

We connect the model manifold, $\mathcal{M}$, to the data space by the residual mapping, $ r:\mathcal{M}\to \mathcal{D}$. Its corresponding pullback and pushforward mappings are labeled $r^*:T^*\mathcal{D}\to T^*\mathcal{M}$ and $r_*:T\mathcal{M}\to T\mathcal{D}$. We summarize the various mappings between tangent and cotangent bundles of $\mathcal{D}$, $\mathcal{M}$ and $\Lambda$ in Figure \ref{fig:scheme}. The metric in the space of residuals is Euclidean, $g^{(\mathcal{D})}= \delta_{ij} dr^i dr^j$, while its pullback to the model manifold is the Fisher information metric, $g^{(\mathcal{M})}=r^{*}g^{(\mathcal{D})}$.

We would like to connect the cotangent bundle of the space of residuals to the tangent bundle $T\Lambda$. We do this by considering the mapping $\flat r_*V_*:T\Lambda \to T^*\mathcal{D}$. We evaluate the effect of this mapping, shown in red in Figure \ref{fig:scheme}, by computing how it acts upon an arbitrary element of $T\Lambda$,  $A=A^a\partial_a\in T\Lambda$
\begin{align}
    \flat r_* V_* A&=\flat\left(r_*\left(A^a V^{\mu}{}_a \partial_\mu\right)\right)\\
    &=A^a V^{\mu}{}_a \flat\left(r_*\left(\partial_\mu\right)\right)\\
    &=A^a V^{\mu}{}_a \partial_\mu r^i\flat\left(\partial_i\right)\\
    &=A^a V^{\mu}{}_a \partial_\mu r^i\delta_{ij} dr^j\\
    &=A^a V^{\mu}{}_a{\color{black}\sum\limits_{i=1}^N}  \partial_\mu r^idr^i.
\end{align}

We now consider small perturbations to the eigenvalue of the largest-uncertainty eigendirection, $e_0$, whose $\sigma_0$ is reduced by a factor $(1+\alpha)$.  The scaling factor $\alpha$ is used to vary the intensity of the model parameters' uncertainty reduction to obtain a desired amount of reduction to the prospective measurements' uncertainty. This goes both ways, since one can then see by how much measurement needs to be improved to have the desired effect on model parameters' uncertainty. 
One can therefore simulate the desired reduction of parameter errors by varying $\alpha$, then estimate the necessary reduction to measurement errors.

Note that $\alpha$ does not need to be a small quantity, as it simply specifies the desired improvement in the precision of that parameter. It can take large values if the experimental design allows for proportionally smaller measurement errors. In other words, $\alpha$ is constrained only by the realistic limits of achievable measurement improvements, not by any mathematical requirement that it must be $\ll1$. 

This results in a sequence of spaces $\Lambda_{\alpha}$, with its corresponding tangent and cotangent bundles.
We first normalize the eigendirections $e_0(\alpha)\in T\Lambda_\alpha$ 

\begin{align}
e_0(\alpha)&=\frac{1}{\sqrt{g^{(\Lambda_\alpha)}(\partial_0,\partial_0)}}\frac{\partial}{\partial V^0}\\
&=\frac{\sigma_0}{1+\alpha}\partial_0,
\end{align}
and then compute the difference \begin{align}\Delta e_0(\alpha)=e_0-e_0(\alpha)= \frac{\alpha}{1+\alpha}\sigma_0\partial_0,\label{eq:de0}\end{align} pushed forward to the space of residuals, \begin{align}\Delta r=\flat r_* V_*\Delta e_0(\alpha).\end{align}
Since we have related changes in residuals to the changes in $e_0$, we drop the normalization and use a modified version of Eq.~\ref{eq:de0}: \begin{align}\Delta \tilde{e}_0(\alpha)= \alpha\sigma_0\partial_0,\label{eq:de0b}\end{align}
so that we can freely vary the strength of the reductions in residuals.

We keep the initial assumption $\Delta r\sim \mathcal{N}(0,1)$ to estimate the impact of this change to the derivatives of the $\chi^2$ estimates
\begin{align}
\frac{\partial}{\partial \sigma^i}\chi^2 &= {\color{black}\sum\limits_{k=1}^N} \frac{(y^k-f^k)^2}{-(\sigma^{k})^3}\delta_{ik}=-\frac{(r^i)^2}{\sigma^i}\\
\frac{\partial^2}{\partial \sigma^i\sigma^j}\chi^2 &= 3\frac{(y^i-f^i)^2}{(\sigma^{i})^4}\delta_{ij}=3\left(\frac{r^i}{\sigma^i}\right)^2\delta_{ij}.
\end{align}
These derivatives are then used to estimate the expected squared change of the residuals, $E[d r\otimes d r]$, in a process similar to the one used to derive the Fisher information metric
\begin{align}
E[d r\otimes d r]
&=E\left[\left(\frac{\partial^2\chi^2}{\partial \sigma^i\sigma^j} d\sigma^i\otimes d\sigma^j\right)\right]\\
&={\color{black}\sum\limits_{i=1}^N\sum\limits_{j=1}^N} 3E\left[\frac{(r^i)^2}{(\sigma^i)^2}\delta_{ij}\right] d\sigma^i \otimes d\sigma^j\\
&={\color{black}\sum\limits_{i=1}^N}3 \left(\frac{d\sigma^i}{\sigma^i}\right)^2,
\end{align}
where we have used the property of the residuals $E[(r^i)^2]=1$.
We connect $E[d r\otimes d r]$ to the numerical perturbation of the eigenvalue by $\alpha$, $E[\Delta r\otimes \Delta r]$ 
\begin{align}
E[\Delta r\otimes \Delta r]&={\color{black}\sum\limits_{i=1}^N\sum\limits_{j=1}^N} E[(\Delta r^i)\otimes (\Delta r^j)]\\
&={\color{black}\sum\limits_{i=1}^N\sum\limits_{j=1}^N}\delta^{ij} E[(\Delta r^i)\otimes (\Delta r^j)]\\
&={\color{black}\sum\limits_{i=1}^N}E\left[(\Delta r^i)^2\right]\\
&={\color{black}\sum\limits_{i=1}^N} \sigma_0^2 \alpha^2 (V_{\mu}^0 g^{(\mathcal{M})}{}^{\mu\nu}\partial_\nu r^i)^2 dr^i\otimes dr^i\end{align}

This connection yields the following update to the relative error $d\sigma^i/\sigma^i$ {\color{black}and the error shifts $d\sigma^i$}
\begin{align}
\left|\frac{d\sigma^i}{\sigma^i}\right|&=\left|\sigma_0 \frac{\alpha}{\sqrt{3}} V_{\mu}^0 g^{(\mathcal{M})}{}^{\mu\nu}\partial_\nu r^i\right|\label{eq:dss}\\
\left|d\sigma^i\right|&=\left|\sigma_0 \frac{\alpha}{\sqrt{3}} V_{\mu}^0 g^{(\mathcal{M})}{}^{\mu\nu}\partial_\nu f^i\right|.\label{eq:ds}
\end{align}
We use properties of the fitted model to inform us which measurements are worth repeating with greater precision.

\section{Fission fragment mass distribution}\label{sec:cgmf_description}
\cgmf{} uses a three-Gaussian representation for the pre-neutron fission fragment mass distribution
\begin{eqnarray}
Y(A;E_n) = G_0(A)+G_1(A)+G_2(A),
\label{eq:YA}
\end{eqnarray}
where $G_0$ corresponds to a symmetric mode, and $G_1$ and $G_2$ correspond to the two asymmetric modes:
\begin{align}
G_0(A) &= \frac{w_0}{\sigma_0\sqrt{2\pi}}{\rm exp}\left( -\frac{(A-\overline{A})^2}{2\sigma_0^2} \right),\label{eq:G012}\\
G_{1,2}(A) &= \frac{w_{1,2}}{\sigma_{1,2}\sqrt{2\pi}}\Big[  {\rm exp}\left( -\frac{(A-\mu_{1,2})^2}{2\sigma_{1,2}^2}\right) \\
&+ {\rm exp}\left( -\frac{\left(A-(A_p-\mu_{1,2})\right)^2}{2\sigma_{1,2}^2}\right)\Big].
\end{align}
Here, $\overline{A}=A_p/2$, with $A_p$ being the mass of the parent fissioning nucleus. In the case where the pre-fission neutrons are emitted, $\overline{A}$ can differ from the original compound nucleus $(Z_c,A_c)$.  The Gaussians require the means, $\mu_i$, the weights, $w_i$, and the widths, $\sigma_i$, of the symmetric Gaussian mode ($i=0$) and the two asymmetric Gaussian modes ($i\in\{1,2\}$). 

The Gaussian mode parameters have an energy dependence, and \cgmf{} models the means and widths as linear in the incident neutron energy
\begin{align}
\mu_i &= \mu_i^{(0)} + \mu_i^{(1)}E_n \\
\sigma_i &= \sigma_i^{(0)} + \sigma_i^{(1)}E_n.
\end{align}
The weights are given by a Fermi function
\begin{eqnarray}
w_i = \frac{1}{1+\exp[(E_n-w_i^{(0)})/w_i^{(1)}]}.
\end{eqnarray}

The following conservation equation, $w_0 = 2 - 2w_1 - 2w_2$, is used to calculate the weight of the symmetric Gaussian $w_0$.
\begin{figure*}
\begin{tikzpicture}[scale=.95, transform shape,block/.style={
      rectangle,
      draw=blue,
      thick,
      fill=blue!20,
      align=center,
      rounded corners,
      minimum height=2cm,scale=1.05
    }]
    \node[block,text width=2cm,align=center] (a) at (0,0)   {Measurement errors $\sigma^i$}; 
    \node[block,text width=3cm,align=center] (b) at (3.5,0)   {$g_{\mu\nu}
    =\sum\limits_{i=1}^N {\partial_\mu}\frac{f^i}{\sigma^i}{\partial_\nu}\frac{f^i}{\sigma^i}$}; 
    \node[block,text width=3cm,align=center] (c) at (7.5,0)   {$\theta_{\text{sim}}\sim\mathcal{N}(\theta_{\text{default}},g^{-1}{}^{(\mathcal{M})})$}; 
    \node[block,text width=3.5cm,align=center] (d) at (11.75,0)   {$g^{\text{sim}}_{\mu\nu}=\sum\limits_{i=1}^N {\partial_\mu}\frac{f^i}{\sigma^i_{sim}}{\partial_\nu}\frac{f^i}{\sigma^i_{sim}}$}; 
    \node[block,text width=4cm,align=center] (e) at (16.5,0)   {$\left|d\sigma^i_{\text{sim}}\right|=\left|\sigma_0 \frac{\alpha}{\sqrt{3}} V_{\mu}^0 g^{(\mathcal{M})}{}^{\mu\nu}_{sim}\partial_\nu f^i\right|$}; 
    \draw[->,thick] (a)--(b);
    \draw[->,thick] (b)--(c);
    \draw[->,thick] (c)--(d);
    \draw[->,thick] (d)--(e);

\end{tikzpicture}  
\caption{Schematic overview of the procedure used to derive Monte Carlo simulations of the measurement error reductions. For each sample, $\theta_{\text{sim}}$, the measurement error reductions, $d\sigma^i_{\text{sim}}$, are computed using $g^{\text{sim}}_{\mu\nu}$.}\label{fig:Schema}
\end{figure*}
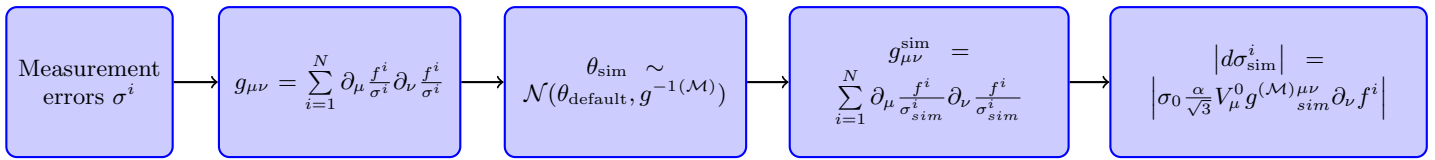
\section{Results\label{sec:results}}
We have chosen to vary a subset of the \cgmf{} parameters ($w_0^{(0)}$, $w_0^{(1)}$, $w_1^{(0)}$, $w_1^{(1)}$, $\mu_0^{(0)}$, $\mu_1^{(0)}$, and $\sigma_0^{(0)}$), for which we numerically estimate the jacobians by shifting the \cgmf{} input parameters by a value of $\pm 0.1$. We use these jacobians to compute $g^{(\mathcal{M})}$ for this parameter subspace.

We have computed the error shifts, $d\sigma^i$, for each atomic mass bin of the \cgmf{} simulations of the {\color{black}mass distributions} of ${}^{252}$Cf for the different relative values of the scaling parameter $\alpha$. 
\subsection{Poissonian errors}\label{sec:poisson}
We have applied the method outlined in section \ref{sec:difgeo2} to the pre-neutron emission mass distributions of the spontaneous fission of ${}^{252}$Cf, using the 10000 \cgmf{} simulations. To do this, we group the outputs of \cgmf{} fission event simulations by atomic mass, count the number of fission events in each atomic mass bin, $N^i$, and compute the  pre-neutron emission mass yields as
\begin{align}
    Y^i = {\color{black}2\frac{N^i}{\sum\limits_{j=1}^N N^j}.}
\end{align}
We estimate the errors {\color{black} for this dataset} using the Poissonian error model
\begin{align}
    \sigma^i={\color{black}2\frac{\sqrt{N^i}}{\sum\limits_{j=1}^N N^j}.}
\end{align}

To study the reliability of these estimates, we run 100 Monte Carlo simulations centered around the default \cgmf{} parameter values\cite{Talou2021}, $\theta_{\text{default}}$, and modeled by a covariance matrix $g^{-1}{}^{(\mathcal{M})}$ as a normal distribution, $\mathcal{N}(\theta_{\text{default}},g^{-1}{}^{(\mathcal{M})})$. We run \cgmf{} for each simulated point in the parameter space and compute the corresponding $\sigma^i_{\text{sim}} = \sigma^i - d\sigma^i$, as in Eq. (\ref{eq:ds}). The procedure is summarized in Fig.~\ref{fig:Schema}.
\begin{figure}[b]
    \centering
\includegraphics[width=1.1\columnwidth]{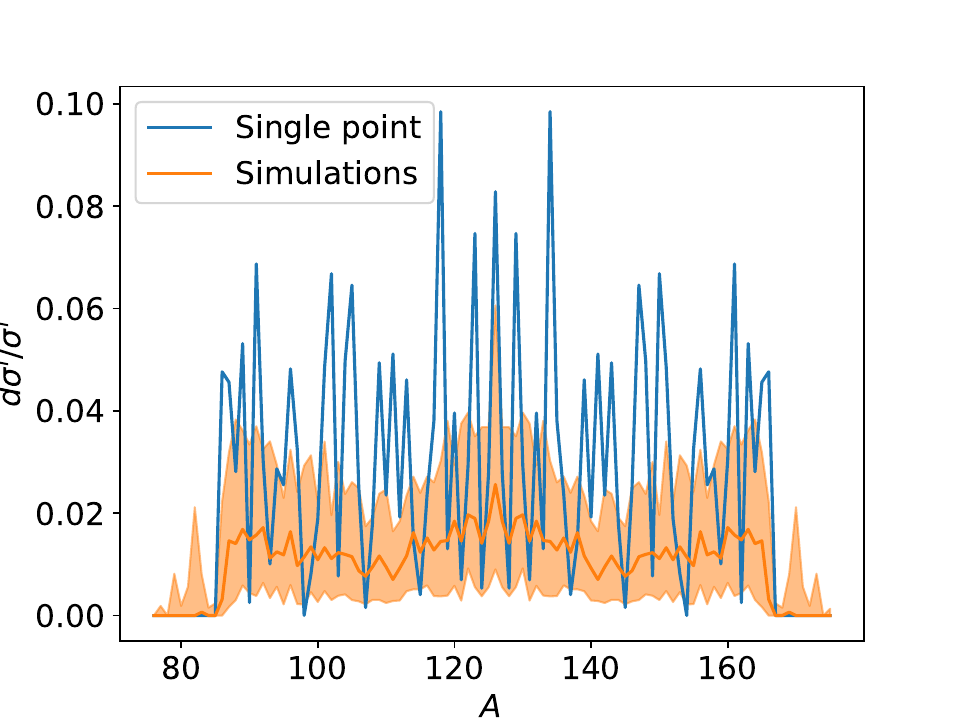}
    \caption{Relative change of the measurement errors (blue) for $\alpha=1$ for Poissonian errors. The orange shaded region is the 16-84 percentile interval computed using Monte Carlo simulations and the orange line is the median.}
    \label{fig:dsig}
    \end{figure}
In Fig. \ref{fig:dsig} we show the results of these simulations as the 16-84 percentile interval for the relative errors $d\sigma^i/\sigma^i$, shown for $\alpha=1$ as an example. As expected the median (orange line) shows less oscillations than the relative error computed only at one point. Even when considering the median over all simulations, Fig.~\ref{fig:dsig}  shows that the contribution of individual atomic mass bins to the relative change  $d\sigma^i/\sigma^i$ remains highly non-uniform.

\begin{figure}[b]
    \centering
    \includegraphics[width=1.1\columnwidth]{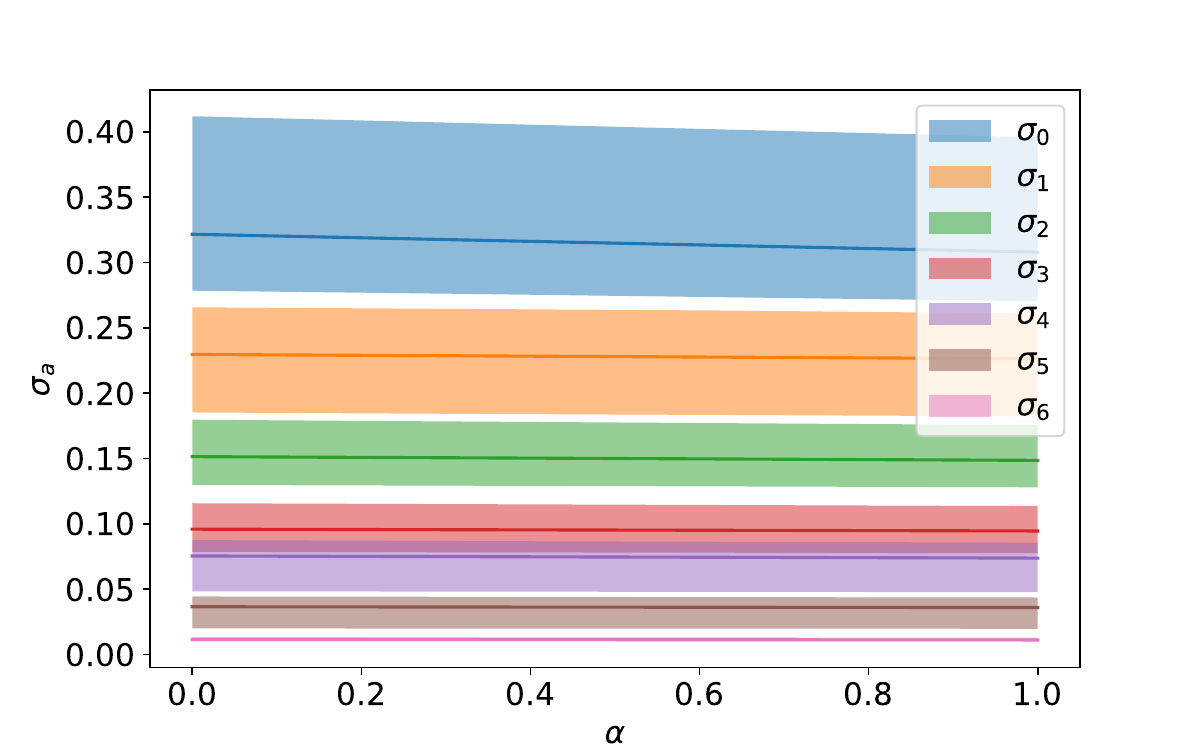}
    \caption{Inverses of the FIM eigenvalues as functions of $\alpha$  for Poissonian errors. Shown are the medians (solid lines) and the 16-84 percentile intervals (shaded regions) of covariance matrix eigenvalues computed using Monte Carlo simulations.}
    \label{fig:Lsig}
    \end{figure}

\begin{figure}[t]
    \centering
    \includegraphics[width=1.\columnwidth]{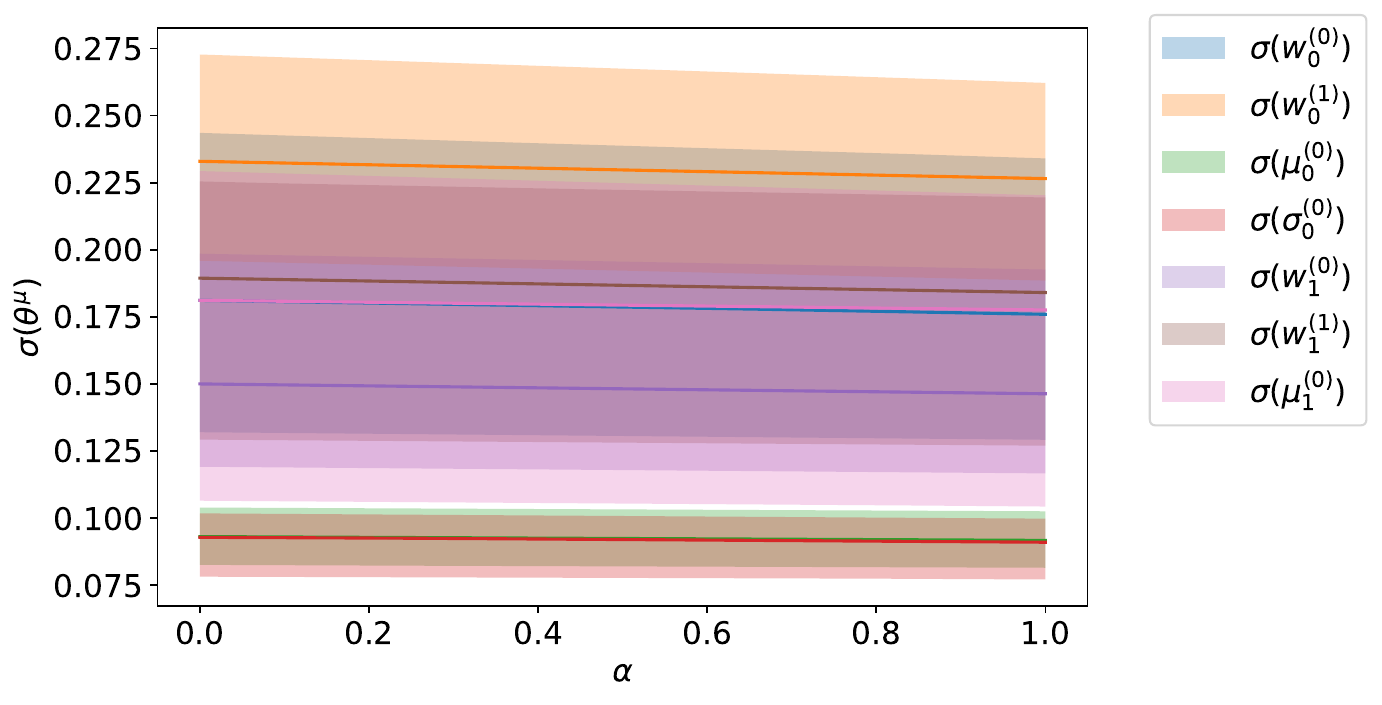}
     \caption{Parameter errors as functions of $\alpha$  for Poissonian errors. Shown are the medians (solid lines) and the 16-84 percentile intervals (shaded regions) of the square roots of the covariance matrix diagonal values computed using Monte Carlo simulations.}
    \label{fig:Msig}
    \end{figure}

\begin{figure}[H]
    \includegraphics[width=1.1\columnwidth]{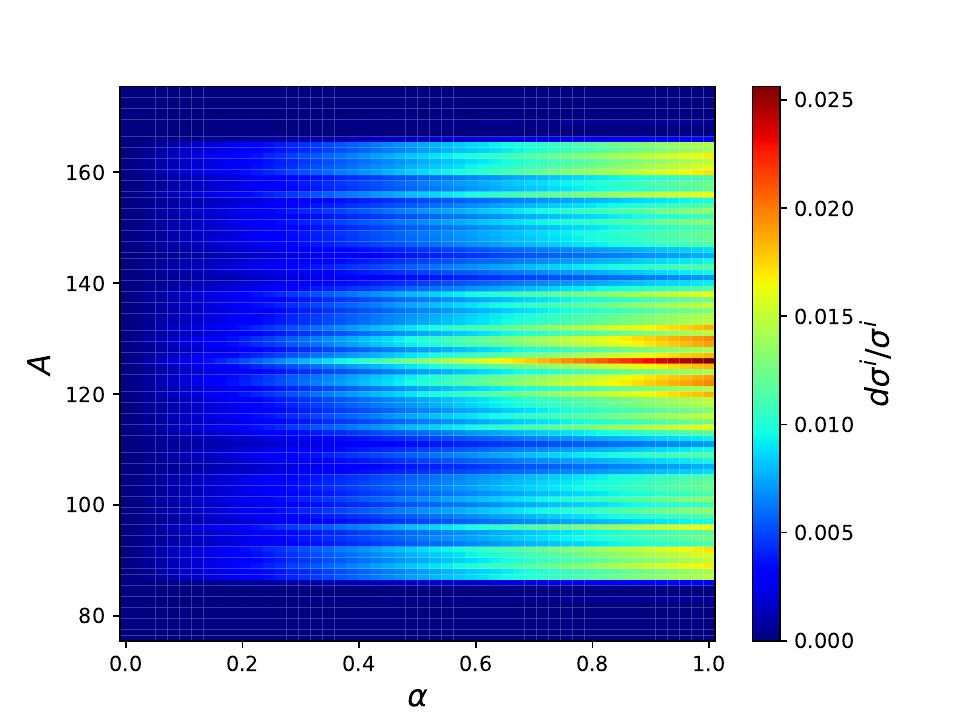}
     \caption{Relative change of the pre-neutron emission mass yields measurement errors as a function of $\alpha$  for Poissonian errors. }
    \label{fig:dsigalpha}
\end{figure}

Figures~\ref{fig:Lsig}--\ref{fig:dsigalpha} summarize the behaviour of the algorithm.
Figure~\ref{fig:Lsig} shows the eigenvalues of the parameter covariance matrix (i.e., the inverses of the FIM eigenvalues) after a single step of error reduction, plotted as a function of the scaling factor~$\alpha$. By construction, the procedure targets the largest-uncertainty eigendirection~$\sigma_{0}$\footnote{The $\sigma_0$ used here is unrelated to the one in Eq.~\ref{eq:G012}.}; consequently, all remaining covariance matrix eigenvalues satisfy $\sigma_{a} \leq \sigma_{0}(\alpha)$. Figure~\ref{fig:Msig} presents the corresponding parameter errors in the model space~$M$ as~$\alpha$ varies, while Fig.~\ref{fig:dsigalpha} shows the relative reductions in the measurement errors~$d\sigma_{i} / \sigma_{i}$ implied by the same~$\alpha$ values.

The median improvement across all Monte Carlo simulations is reported in these figures, allowing a direct ``lookup'' usage: for a desired reduction in parameter uncertainties (read from Fig.~\ref{fig:Msig}), one can determine the necessary measurement precision improvement (read from Fig.~\ref{fig:dsigalpha}). Table~\ref{tab:Shifttab} lists the ten atomic mass bins with the largest relative error shifts for representative~$\alpha$ values, indicating the most effective measurements to repeat. As expected, increasing~$\alpha$ yields progressively smaller parameter errors, but this comes at the cost of proportionally larger reductions in the corresponding measurement uncertainties.

The asymmetry of the uncertainties in Table~\ref{tab:Shifttab} arises from our choice to report the 16th--84th percentile interval from the Monte Carlo distributions, rather than a standard deviation around the median. This approach captures the actual shape of the distributions of $d\sigma_i / \sigma_i$, which are not necessarily symmetric due to the nonlinear mapping from parameter space to measurement-error space. In such cases, percentile-based intervals provide a more robust, non-parametric characterization of uncertainty, especially when the distributions exhibit skewness or extended tails. Importantly, this method does not alter the underlying distributions, which remain essentially unchanged under this form of analysis.

\begin{table}[t]
     \centering
    \begin{tabular}{cccc}
\toprule
$A$ &     $\frac{d\sigma^i}{\sigma^i}(\alpha=0.1)$&     $\frac{d\sigma^i}{\sigma^i}(\alpha=0.5)$ &     $\frac{d\sigma^i}{\sigma^i}(\alpha=1)$ \\
\midrule
126 & $0.003\,_{-0.002}^{+0.003}$ & $0.013\,_{-0.008}^{+0.017}$ & $0.026\,_{-0.017}^{+0.035}$ \\
122 & $0.002\,_{-0.001}^{+0.002}$ & $0.010\,_{-0.005}^{+0.010}$ & $0.020\,_{-0.010}^{+0.020}$ \\
130 & $0.002\,_{-0.001}^{+0.002}$ & $0.010\,_{-0.005}^{+0.010}$ & $0.020\,_{-0.010}^{+0.020}$ \\
129 & $0.002\,_{-0.001}^{+0.002}$ & $0.009\,_{-0.007}^{+0.008}$ & $0.019\,_{-0.013}^{+0.016}$ \\
123 & $0.002\,_{-0.001}^{+0.002}$ & $0.009\,_{-0.007}^{+0.008}$ & $0.019\,_{-0.013}^{+0.016}$ \\
120 & $0.002\,_{-0.001}^{+0.001}$ & $0.009\,_{-0.006}^{+0.005}$ & $0.018\,_{-0.013}^{+0.011}$ \\
132 & $0.002\,_{-0.001}^{+0.001}$ & $0.009\,_{-0.006}^{+0.005}$ & $0.018\,_{-0.013}^{+0.011}$ \\
125 & $0.002\,_{-0.001}^{+0.002}$ & $0.009\,_{-0.006}^{+0.009}$ & $0.018\,_{-0.013}^{+0.018}$ \\
127 & $0.002\,_{-0.001}^{+0.002}$ & $0.009\,_{-0.006}^{+0.009}$ & $0.018\,_{-0.013}^{+0.018}$ \\
160 & $0.002\,_{-0.001}^{+0.002}$ & $0.009\,_{-0.005}^{+0.008}$ & $0.017\,_{-0.011}^{+0.015}$ \\
\bottomrule
\end{tabular}  \caption{The mass numbers with the 10 highest median relative error shifts for different values of $\alpha$  for Poissonian errors. The asymmetric errors are computed from the 16-th and 84-th percentiles.}
    \label{tab:Shifttab}
\end{table}
\subsection{Real measurement errors}\label{sec:realres}
\begin{figure}[H]
    \centering
\includegraphics[width=1.1\columnwidth]{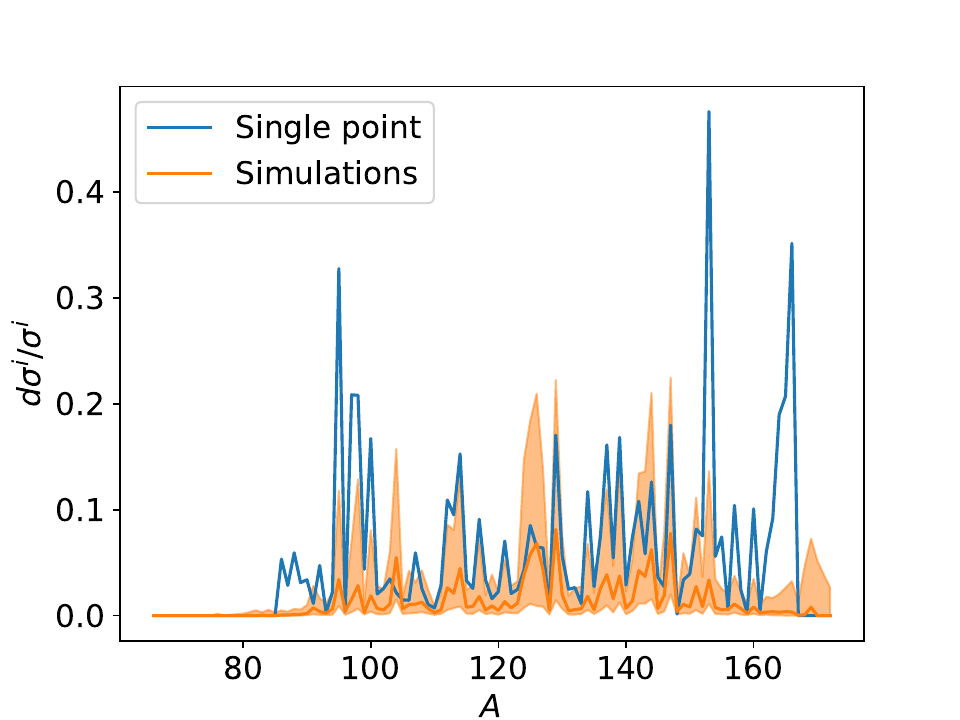}
    \caption{Relative change of the measurement errors (blue) for $\alpha=1$ for measurements of ${}^{252}$Cf. The orange shaded region is the 16-84 percentile interval computed using Monte Carlo simulations and the orange line is the median.}
    \label{fig:dsigm}
    \end{figure}

\begin{figure}[H]
    \centering
    \includegraphics[width=1.1\columnwidth]{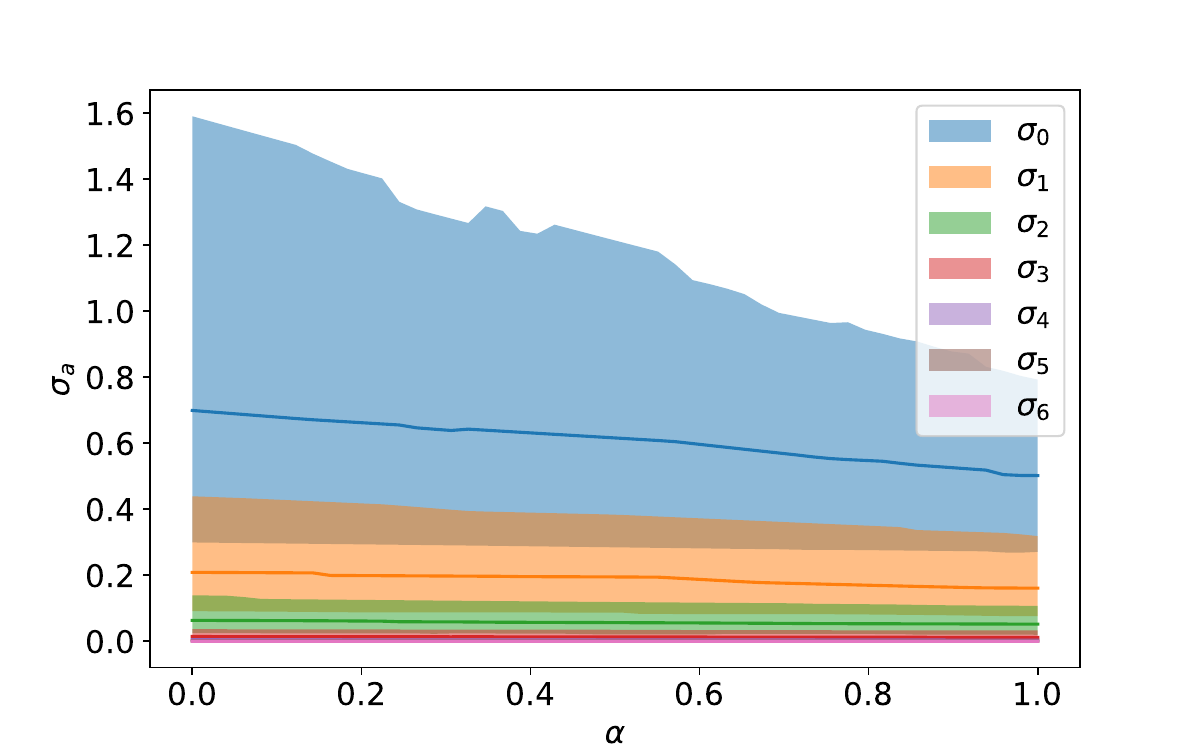}
    \caption{Inverses of the FIM eigenvalues as functions of $\alpha$ for measurements of ${}^{252}$Cf. Shown are the medians (solid lines) and the 16-84 percentile intervals (shaded regions) of covariance matrix eigenvalues computed using Monte Carlo simulations.}
    \label{fig:Lsigm}
     \end{figure}

\begin{figure}[H]
    \centering
    \includegraphics[width=1.\columnwidth]{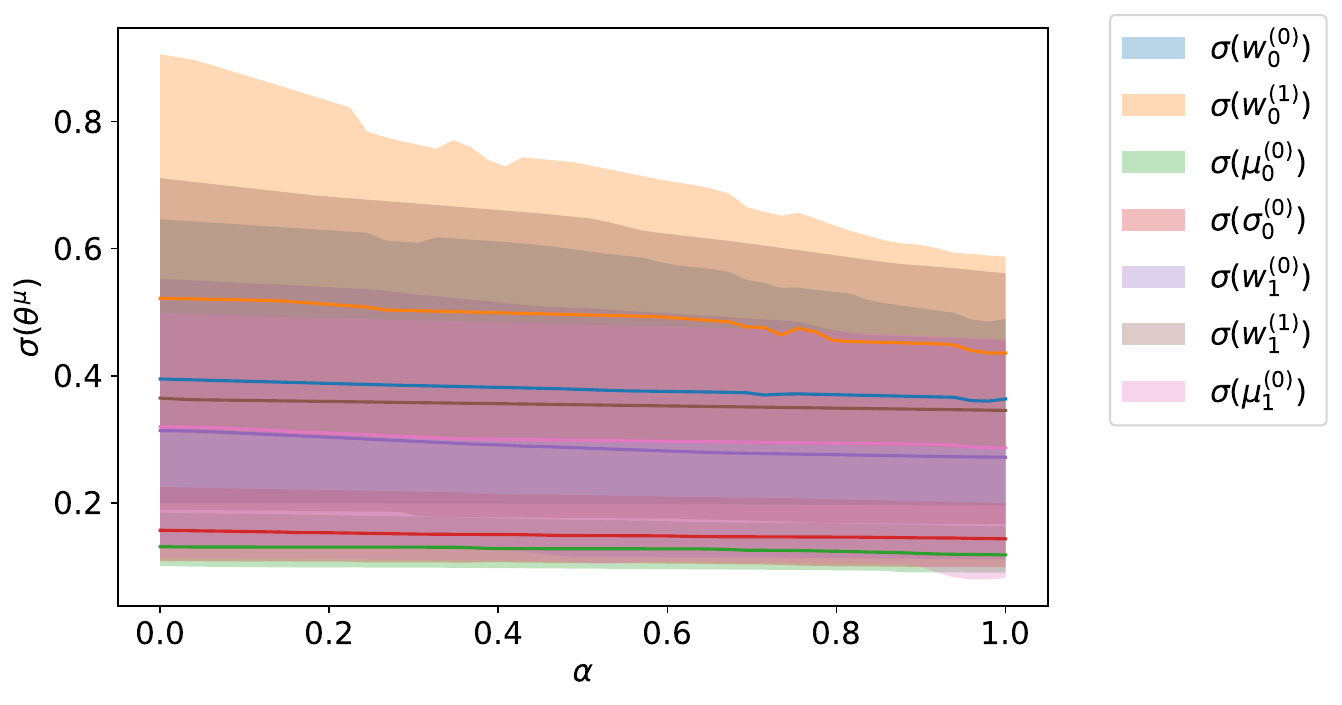}
     \caption{Parameter errors as functions of $\alpha$ for measurements of ${}^{252}$Cf. Shown are the medians (solid lines) and the 16-84 percentile intervals (shaded regions) of the square roots of the covariance matrix diagonal values computed using Monte Carlo simulations.}
    \label{fig:Msigm}
    \end{figure}

\begin{figure}[H]
    \includegraphics[width=1.1\columnwidth]{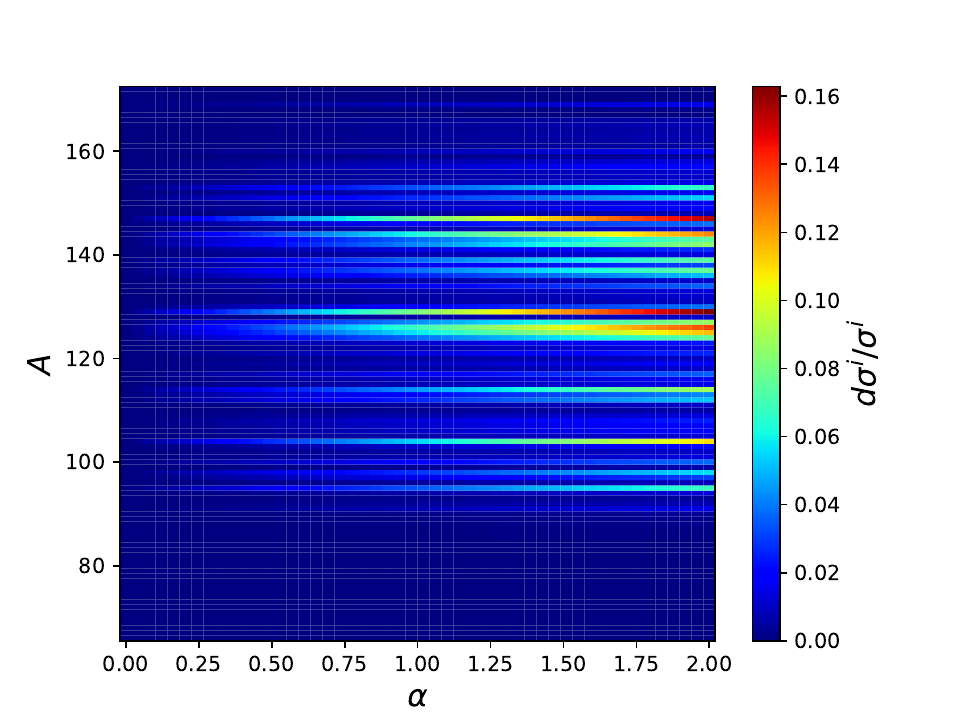}
     \caption{Relative change of the pre-neutron emission mass yields measurement errors for measurements of ${}^{252}$Cf as a function of $\alpha$. }
    \label{fig:dsigalpham}
\end{figure}

\begin{figure}[H]
\includegraphics[width=\columnwidth]{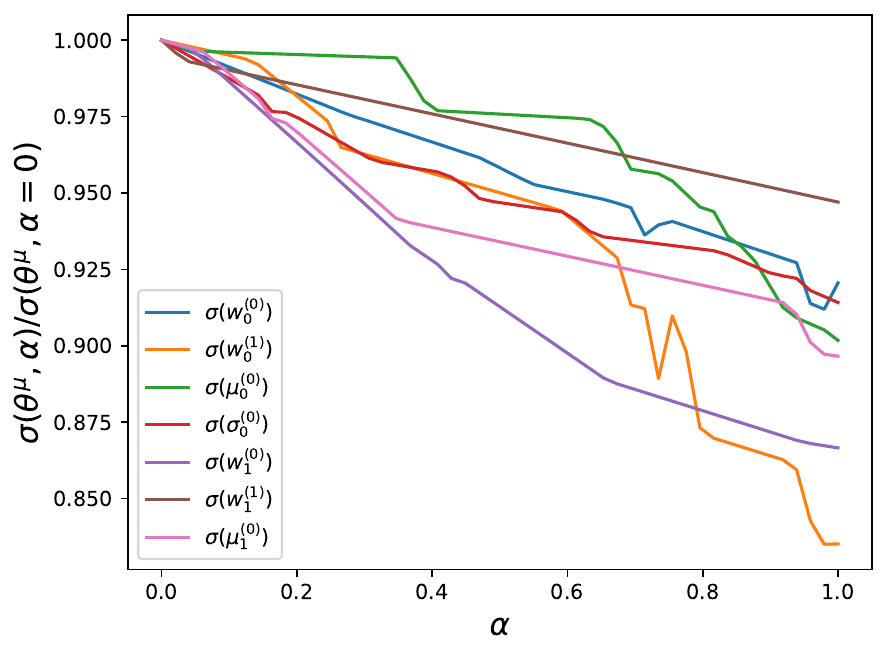}
     \caption{Parameter errors as functions of $\alpha$, scaled using their respective values at $\alpha=0$ for measurements of ${}^{252}$Cf. Shown are the medians  of the square roots of the covariance matrix diagonal values computed using Monte Carlo simulations.}
    \label{fig:Rsigm}
    \end{figure}

 We have also applied our method to the measurements of the ${}^{252}$Cf pre-neutron emission mass yields taken from EXFOR \cite{Gook2014}. Since the only required experimental data for our method are measurement errors, we can compare not only the application to the default \cgmf{} parameter point, but we can also rescale the Monte Carlo sample of the \cgmf{} Jacobians to the real measurement errors, instead of the Poissonian error model. The results of this procedure are shown in Figs. \ref{fig:dsigm}, \ref{fig:Lsigm}, \ref{fig:Msigm}, and \ref{fig:dsigalpham} corresponding to the similar Figs. \ref{fig:dsig}, \ref{fig:Lsig}, \ref{fig:dsigalpha}, and \ref{fig:Msig} for Poissonian errors. We see the same kind of behavior as the toy problem discussed in Sec. \ref{sec:poisson}, albeit with peaks of varying amplitude. For Poissonian errors in Fig.~\ref{fig:dsig}, the results of simulations (orange) generally follow the peaks at the default point (blue), but their peaks are $\sim 5$ times higher than the simulations compared. This is similar to the peaks in real ${}^{252}$Cf uncertainties. Real changes of measurement errors are overall $\sim 10$ times higher in Fig.~\ref{fig:dsigm} than for Poissonian case in Fig.~\ref{fig:dsig}. The inverse FIM eigenvalues (Fig. \ref{fig:Lsigm}) and parameter errors (Fig. \ref{fig:Msigm}) are $\sim 5-10$ times higher than in the toy problem in Figs. \ref{fig:Lsig} and \ref{fig:Msig}. We list the relative errors for  the 10 highest relative error shifts for different values of $\alpha$ in table \ref{tab:Shifttabmeas}. The chosen values of $\alpha$ in the table are different from those in table \ref{tab:Shifttab} because of the different scale of the relative measurement errors in these two datasets. From the parameter errors normalized to their values at $\alpha=0$ (i.e., the case of no reduction) in Fig.~\ref{fig:Rsigm}, we conclude that by reducing measurement uncertainties one can achieve a maximum reduction of up to $\sim15\%$ for the cases of $w_0^{(1)}$ and $w_1^{(0)}$ (orange and purple lines in Fig. \ref{fig:Rsigm}).

\begin{table}[H]
     \centering
       
    \begin{tabular}{cccc}
\toprule
$A$ &     $\frac{d\sigma^i}{\sigma^i}(\alpha=0.1)$&     $\frac{d\sigma^i}{\sigma^i}(\alpha=0.5)$ &     $\frac{d\sigma^i}{\sigma^i}(\alpha=1.0)$ \\
\midrule
129 & $0.008\,_{-0.007}^{+0.014}$ & $0.041\,_{-0.033}^{+0.070}$ & $0.081\,_{-0.067}^{+0.141}$ \\
147 & $0.008\,_{-0.006}^{+0.015}$ & $0.039\,_{-0.029}^{+0.074}$ & $0.078\,_{-0.058}^{+0.147}$ \\
126 & $0.007\,_{-0.006}^{+0.014}$ & $0.034\,_{-0.029}^{+0.071}$ & $0.068\,_{-0.059}^{+0.141}$ \\
144 & $0.006\,_{-0.005}^{+0.015}$ & $0.031\,_{-0.023}^{+0.074}$ & $0.062\,_{-0.047}^{+0.148}$ \\
125 & $0.006\,_{-0.005}^{+0.013}$ & $0.029\,_{-0.023}^{+0.063}$ & $0.058\,_{-0.047}^{+0.126}$ \\
104 & $0.005\,_{-0.004}^{+0.010}$ & $0.027\,_{-0.020}^{+0.051}$ & $0.055\,_{-0.039}^{+0.103}$ \\
114 & $0.004\,_{-0.004}^{+0.009}$ & $0.022\,_{-0.018}^{+0.044}$ & $0.045\,_{-0.036}^{+0.087}$ \\
127 & $0.004\,_{-0.003}^{+0.009}$ & $0.022\,_{-0.017}^{+0.047}$ & $0.043\,_{-0.035}^{+0.095}$ \\
142 & $0.004\,_{-0.003}^{+0.009}$ & $0.021\,_{-0.016}^{+0.046}$ & $0.042\,_{-0.031}^{+0.092}$ \\
137 & $0.004\,_{-0.003}^{+0.008}$ & $0.019\,_{-0.014}^{+0.040}$ & $0.039\,_{-0.029}^{+0.081}$ \\
\bottomrule
\end{tabular} \caption{The atomic numbers with the 10 highest median relative error shifts for different values of $\alpha$ for measurements of ${}^{252}$Cf. The asymmetric errors are computed from the 16-th and 84-th percentiles.}
    \label{tab:Shifttabmeas}
\end{table}

\section{Conclusion\label{sec:conclusion}}
The \cgmf{} code plays a pivotal role in modeling the de-excitation of fission fragments. Operating on an event-by-event basis, this code predicts distributions and correlations of neutrons, photons, and fission fragments. It relies on the Hauser-Feshbach statistical model of nuclear reactions. Currently, the \cgmf{} code focuses on binary fission, accounting for pre-scission neutrons and multi-chance fission.


In this work, we applied information-geometric methods to investigate the impact of measurement errors in the \cgmf{} model for the pre-neutron emission mass distribution model. We demonstrated our method for spontaneous fission of $^{252}$Cf on both a toy problem using a Poissonian error model and to experimental measurement from G\"o\"ok et al., 2014 \cite{Gook2014}. We relate the desired reduction in \cgmf{} parameter errors to the necessary reduction in relative measurement errors to achieve it. We do this by varying the reduction in the largest-uncertainty eigendirection of the \cgmf{} parameter covariance matrix by a scaling parameter $\alpha$ and project this to both the \cgmf{} parameter uncertainty space and to the space of relative measurement errors. This is essentially a lookup table that connects measurement uncertainties to parameter uncertainties. We found the atomic numbers at which we expect that the increase in experimental accuracy of pre-neutron emission mass yields would yield the greatest improvement in the accuracy of \cgmf{} model parameters.

Beyond the specific application to CGMF and nuclear fission modeling, the information-geometric framework developed in this work represents a novel and broadly applicable approach. Its formulation is not limited to nuclear data analysis, and can be directly adapted to a wide range of disciplines where the interplay between measurement uncertainties and parameter constraints plays a decisive role.

\acknowledgments
This work was performed under the auspice of the U.S. Department of Energy by Los Alamos National Laboratory under Contract 89233218CNA000001. Research reported in this publication was supported by the U.S. Department of Energy LDRD program (Project No. 20240004DR and Project No. 20240031DR) at Los Alamos National Laboratory.

\appendix
\section{Musical isomorphisms\label{sec:musiciso}}
The raising of indices is performed by the sharp operator $\#: T^*\Lambda\to T\Lambda$. This operation transforms a covariant vector with lower indices, $A=A_a dV^a\in T^*\Lambda$, to a contravariant vector with upper indices, $\# A=A^a \partial_a=(A_b g^{ab})\partial_a$. Its inverse is the flat operator, $\flat: T\Lambda\to T^*\Lambda$, which corresponds to lowering vector indices, $A_a=g_{ab}A^b$. These operators can also act on product spaces, e.g. on the metric
    \begin{align}
    \# g^{(\Lambda)}&=\left[ \# g^{(\Lambda)}\right]^{ab}\partial_a\otimes \partial_b=g^{-1}{}^{(\Lambda)}\\
    \flat g^{-1}{}^{(\Lambda)}&=\left[\flat g^{-1}{}^{(\Lambda)}\right]_{ab}dV^a\otimes dV^b=g^{(\Lambda)}.
\end{align}
    If the spaces $\Lambda$ and $\mathcal{M}$ are connected by a mapping $V^T:\mathcal{M}\to\Lambda$, the tangent bundle mapping is performed by the pushforward mapping, $V^T_*:T\mathcal{M}\to T\Lambda$. The cotangent bundles are connected by the pullback mapping, $V^{T*}:T^*\Lambda\to T^*\mathcal{M}$. These operations have a direct generalization to product spaces. The pullback operator $V^{T*}$ transforms the metric $g^{(\Lambda)}$ to the metric $g^{(\mathcal{M})}$
    \begin{align}
    V^{T*}g^{(\Lambda)}&={\color{black}\sum\limits_{a=1}^M}\lambda_a (V_{\mu}{}^a d\theta^\mu)\otimes(V_{\nu}{}^a d\theta^\nu)\\
    &=[Vg^{(\Lambda)}V^T]_{\mu\nu} d\theta^\mu\otimes d\theta^\nu\\
    &=g^{(\mathcal{M})}.
    \end{align}
    The inverse metric inverse, however, is transformed by the pushforward operator $V_*^T$ from $g^{-1}{}^{(\mathcal{M})}$ to $g^{-1}{}^{(\Lambda)}$
    \begin{align}
    V_*^T(g^{-1}{}^{(\mathcal{M})})&=V_*^T(g^{(\mathcal{M})}{}^{\mu\nu}\partial_\mu\otimes\partial_\nu)\\
    &=g^{(\mathcal{M})}{}^{\mu\nu}V_{\mu}{}^a V_{\nu}{}^b\frac{\partial}{\partial V^a}\otimes \frac{\partial}{\partial V^b}\\
    &={\color{black}\sum\limits_{a=1}^M} \sigma_a^2 \partial_a\otimes \partial_a\\
    &=g^{-1}{}^{(\Lambda)}.
    \end{align}
    We therefore need the inverse transform to transform  $g^{-1}{}^{(\Lambda)}$ into $g^{-1}{}^{(\mathcal{M})}$, i.e. $g^{-1}{}^{(\mathcal{M})}=V_* g^{-1}{}^{(\Lambda)}$.

\bibliography{references.bib}

\begin{thebibliography}{50}%
\makeatletter
\providecommand \@ifxundefined [1]{%
 \@ifx{#1\undefined}
}%
\providecommand \@ifnum [1]{%
 \ifnum #1\expandafter \@firstoftwo
 \else \expandafter \@secondoftwo
 \fi
}%
\providecommand \@ifx [1]{%
 \ifx #1\expandafter \@firstoftwo
 \else \expandafter \@secondoftwo
 \fi
}%
\providecommand \natexlab [1]{#1}%
\providecommand \enquote  [1]{``#1''}%
\providecommand \bibnamefont  [1]{#1}%
\providecommand \bibfnamefont [1]{#1}%
\providecommand \citenamefont [1]{#1}%
\providecommand \href@noop [0]{\@secondoftwo}%
\providecommand \href [0]{\begingroup \@sanitize@url \@href}%
\providecommand \@href[1]{\@@startlink{#1}\@@href}%
\providecommand \@@href[1]{\endgroup#1\@@endlink}%
\providecommand \@sanitize@url [0]{\catcode `\\12\catcode `\$12\catcode
  `\&12\catcode `\#12\catcode `\^12\catcode `\_12\catcode `\%12\relax}%
\providecommand \@@startlink[1]{}%
\providecommand \@@endlink[0]{}%
\providecommand \url  [0]{\begingroup\@sanitize@url \@url }%
\providecommand \@url [1]{\endgroup\@href {#1}{\urlprefix }}%
\providecommand \urlprefix  [0]{URL }%
\providecommand \Eprint [0]{\href }%
\providecommand \doibase [0]{https://doi.org/}%
\providecommand \selectlanguage [0]{\@gobble}%
\providecommand \bibinfo  [0]{\@secondoftwo}%
\providecommand \bibfield  [0]{\@secondoftwo}%
\providecommand \translation [1]{[#1]}%
\providecommand \BibitemOpen [0]{}%
\providecommand \bibitemStop [0]{}%
\providecommand \bibitemNoStop [0]{.\EOS\space}%
\providecommand \EOS [0]{\spacefactor3000\relax}%
\providecommand \BibitemShut  [1]{\csname bibitem#1\endcsname}%
\let\auto@bib@innerbib\@empty
\bibitem [{\citenamefont {{Randrup}}\ and\ \citenamefont
  {{M{\"o}ller}}(2011)}]{Randrup:2011}%
  \BibitemOpen
  \bibfield  {author} {\bibinfo {author} {\bibfnamefont {J.}~\bibnamefont
  {{Randrup}}}\ and\ \bibinfo {author} {\bibfnamefont {P.}~\bibnamefont
  {{M{\"o}ller}}},\ }\href {https://doi.org/10.1103/PhysRevLett.106.132503}
  {\bibfield  {journal} {\bibinfo  {journal} {\prl}\ }\textbf {\bibinfo
  {volume} {106}},\ \bibinfo {eid} {132503} (\bibinfo {year} {2011})},\ \Eprint
  {https://arxiv.org/abs/1103.0535} {arXiv:1103.0535 [nucl-th]} \BibitemShut
  {NoStop}%
\bibitem [{\citenamefont {{M{\"o}ller}}\ and\ \citenamefont
  {{Ichikawa}}(2015)}]{Moller:2015}%
  \BibitemOpen
  \bibfield  {author} {\bibinfo {author} {\bibfnamefont {P.}~\bibnamefont
  {{M{\"o}ller}}}\ and\ \bibinfo {author} {\bibfnamefont {T.}~\bibnamefont
  {{Ichikawa}}},\ }\href {https://doi.org/10.1140/epja/i2015-15173-1}
  {\bibfield  {journal} {\bibinfo  {journal} {European Physical Journal A}\
  }\textbf {\bibinfo {volume} {51}},\ \bibinfo {eid} {173} (\bibinfo {year}
  {2015})},\ \Eprint {https://arxiv.org/abs/1508.05964} {arXiv:1508.05964
  [nucl-th]} \BibitemShut {NoStop}%
\bibitem [{\citenamefont {{Mumpower}}\ \emph {et~al.}(2020)\citenamefont
  {{Mumpower}}, \citenamefont {{Jaffke}}, \citenamefont {{Verriere}},\ and\
  \citenamefont {{Randrup}}}]{Mumpower:2020}%
  \BibitemOpen
  \bibfield  {author} {\bibinfo {author} {\bibfnamefont {M.~R.}\ \bibnamefont
  {{Mumpower}}}, \bibinfo {author} {\bibfnamefont {P.}~\bibnamefont
  {{Jaffke}}}, \bibinfo {author} {\bibfnamefont {M.}~\bibnamefont
  {{Verriere}}},\ and\ \bibinfo {author} {\bibfnamefont {J.}~\bibnamefont
  {{Randrup}}},\ }\href {https://doi.org/10.1103/PhysRevC.101.054607}
  {\bibfield  {journal} {\bibinfo  {journal} {\prc}\ }\textbf {\bibinfo
  {volume} {101}},\ \bibinfo {eid} {054607} (\bibinfo {year} {2020})},\ \Eprint
  {https://arxiv.org/abs/1911.06344} {arXiv:1911.06344 [nucl-th]} \BibitemShut
  {NoStop}%
\bibitem [{\citenamefont {{Schunck}}\ \emph {et~al.}(2014)\citenamefont
  {{Schunck}}, \citenamefont {{Duke}}, \citenamefont {{Carr}},\ and\
  \citenamefont {{Knoll}}}]{Schunck:2014}%
  \BibitemOpen
  \bibfield  {author} {\bibinfo {author} {\bibfnamefont {N.}~\bibnamefont
  {{Schunck}}}, \bibinfo {author} {\bibfnamefont {D.}~\bibnamefont {{Duke}}},
  \bibinfo {author} {\bibfnamefont {H.}~\bibnamefont {{Carr}}},\ and\ \bibinfo
  {author} {\bibfnamefont {A.}~\bibnamefont {{Knoll}}},\ }\href
  {https://doi.org/10.1103/PhysRevC.90.054305} {\bibfield  {journal} {\bibinfo
  {journal} {\prc}\ }\textbf {\bibinfo {volume} {90}},\ \bibinfo {eid} {054305}
  (\bibinfo {year} {2014})},\ \Eprint {https://arxiv.org/abs/1311.2616}
  {arXiv:1311.2616 [nucl-th]} \BibitemShut {NoStop}%
\bibitem [{\citenamefont {{Bulgac}}\ \emph {et~al.}(2016)\citenamefont
  {{Bulgac}}, \citenamefont {{Magierski}}, \citenamefont {{Roche}},\ and\
  \citenamefont {{Stetcu}}}]{Bulgac:2016}%
  \BibitemOpen
  \bibfield  {author} {\bibinfo {author} {\bibfnamefont {A.}~\bibnamefont
  {{Bulgac}}}, \bibinfo {author} {\bibfnamefont {P.}~\bibnamefont
  {{Magierski}}}, \bibinfo {author} {\bibfnamefont {K.~J.}\ \bibnamefont
  {{Roche}}},\ and\ \bibinfo {author} {\bibfnamefont {I.}~\bibnamefont
  {{Stetcu}}},\ }\href {https://doi.org/10.1103/PhysRevLett.116.122504}
  {\bibfield  {journal} {\bibinfo  {journal} {\prl}\ }\textbf {\bibinfo
  {volume} {116}},\ \bibinfo {eid} {122504} (\bibinfo {year} {2016})},\ \Eprint
  {https://arxiv.org/abs/1511.00738} {arXiv:1511.00738 [nucl-th]} \BibitemShut
  {NoStop}%
\bibitem [{\citenamefont {{Regnier}}\ \emph {et~al.}(2016)\citenamefont
  {{Regnier}}, \citenamefont {{Dubray}}, \citenamefont {{Schunck}},\ and\
  \citenamefont {{Verri{\`e}re}}}]{Regnier:2016}%
  \BibitemOpen
  \bibfield  {author} {\bibinfo {author} {\bibfnamefont {D.}~\bibnamefont
  {{Regnier}}}, \bibinfo {author} {\bibfnamefont {N.}~\bibnamefont {{Dubray}}},
  \bibinfo {author} {\bibfnamefont {N.}~\bibnamefont {{Schunck}}},\ and\
  \bibinfo {author} {\bibfnamefont {M.}~\bibnamefont {{Verri{\`e}re}}},\ }\href
  {https://doi.org/10.1103/PhysRevC.93.054611} {\bibfield  {journal} {\bibinfo
  {journal} {\prc}\ }\textbf {\bibinfo {volume} {93}},\ \bibinfo {eid} {054611}
  (\bibinfo {year} {2016})},\ \Eprint {https://arxiv.org/abs/1603.08824}
  {arXiv:1603.08824 [nucl-th]} \BibitemShut {NoStop}%
\bibitem [{\citenamefont {{Sierk}}(2017)}]{Sierk:2017}%
  \BibitemOpen
  \bibfield  {author} {\bibinfo {author} {\bibfnamefont {A.~J.}\ \bibnamefont
  {{Sierk}}},\ }\href {https://doi.org/10.1103/PhysRevC.96.034603} {\bibfield
  {journal} {\bibinfo  {journal} {\prc}\ }\textbf {\bibinfo {volume} {96}},\
  \bibinfo {eid} {034603} (\bibinfo {year} {2017})}\BibitemShut {NoStop}%
\bibitem [{\citenamefont {{Usang}}\ \emph {et~al.}(2017)\citenamefont
  {{Usang}}, \citenamefont {{Ivanyuk}}, \citenamefont {{Ishizuka}},\ and\
  \citenamefont {{Chiba}}}]{Usang:2017}%
  \BibitemOpen
  \bibfield  {author} {\bibinfo {author} {\bibfnamefont {M.~D.}\ \bibnamefont
  {{Usang}}}, \bibinfo {author} {\bibfnamefont {F.~A.}\ \bibnamefont
  {{Ivanyuk}}}, \bibinfo {author} {\bibfnamefont {C.}~\bibnamefont
  {{Ishizuka}}},\ and\ \bibinfo {author} {\bibfnamefont {S.}~\bibnamefont
  {{Chiba}}},\ }\href {https://doi.org/10.1103/PhysRevC.96.064617} {\bibfield
  {journal} {\bibinfo  {journal} {\prc}\ }\textbf {\bibinfo {volume} {96}},\
  \bibinfo {eid} {064617} (\bibinfo {year} {2017})},\ \Eprint
  {https://arxiv.org/abs/1712.05616} {arXiv:1712.05616 [nucl-th]} \BibitemShut
  {NoStop}%
\bibitem [{\citenamefont {{Regnier}}\ \emph {et~al.}(2019)\citenamefont
  {{Regnier}}, \citenamefont {{Dubray}},\ and\ \citenamefont
  {{Schunck}}}]{Regnier:2019}%
  \BibitemOpen
  \bibfield  {author} {\bibinfo {author} {\bibfnamefont {D.}~\bibnamefont
  {{Regnier}}}, \bibinfo {author} {\bibfnamefont {N.}~\bibnamefont
  {{Dubray}}},\ and\ \bibinfo {author} {\bibfnamefont {N.}~\bibnamefont
  {{Schunck}}},\ }\href {https://doi.org/10.1103/PhysRevC.99.024611} {\bibfield
   {journal} {\bibinfo  {journal} {\prc}\ }\textbf {\bibinfo {volume} {99}},\
  \bibinfo {eid} {024611} (\bibinfo {year} {2019})},\ \Eprint
  {https://arxiv.org/abs/1810.08402} {arXiv:1810.08402 [nucl-th]} \BibitemShut
  {NoStop}%
\bibitem [{\citenamefont {{Talou}}\ \emph {et~al.}(2021)\citenamefont
  {{Talou}}, \citenamefont {{Stetcu}}, \citenamefont {{Jaffke}}, \citenamefont
  {{Rising}}, \citenamefont {{Lovell}},\ and\ \citenamefont
  {{Kawano}}}]{Talou2021}%
  \BibitemOpen
  \bibfield  {author} {\bibinfo {author} {\bibfnamefont {P.}~\bibnamefont
  {{Talou}}}, \bibinfo {author} {\bibfnamefont {I.}~\bibnamefont {{Stetcu}}},
  \bibinfo {author} {\bibfnamefont {P.}~\bibnamefont {{Jaffke}}}, \bibinfo
  {author} {\bibfnamefont {M.~E.}\ \bibnamefont {{Rising}}}, \bibinfo {author}
  {\bibfnamefont {A.~E.}\ \bibnamefont {{Lovell}}},\ and\ \bibinfo {author}
  {\bibfnamefont {T.}~\bibnamefont {{Kawano}}},\ }\href
  {https://doi.org/10.1016/j.cpc.2021.108087} {\bibfield  {journal} {\bibinfo
  {journal} {Computer Physics Communications}\ }\textbf {\bibinfo {volume}
  {269}},\ \bibinfo {eid} {108087} (\bibinfo {year} {2021})},\ \Eprint
  {https://arxiv.org/abs/2011.10444} {arXiv:2011.10444 [nucl-th]} \BibitemShut
  {NoStop}%
\bibitem [{\citenamefont {{Hauser}}\ and\ \citenamefont
  {{Feshbach}}(1952)}]{Hauser1952}%
  \BibitemOpen
  \bibfield  {author} {\bibinfo {author} {\bibfnamefont {W.}~\bibnamefont
  {{Hauser}}}\ and\ \bibinfo {author} {\bibfnamefont {H.}~\bibnamefont
  {{Feshbach}}},\ }\href {https://doi.org/10.1103/PhysRev.87.366} {\bibfield
  {journal} {\bibinfo  {journal} {Physical Review}\ }\textbf {\bibinfo {volume}
  {87}},\ \bibinfo {pages} {366} (\bibinfo {year} {1952})}\BibitemShut
  {NoStop}%
\bibitem [{\citenamefont {Verbeke}\ \emph {et~al.}(2015)\citenamefont
  {Verbeke}, \citenamefont {Randrup},\ and\ \citenamefont {Vogt}}]{FREYA1}%
  \BibitemOpen
  \bibfield  {author} {\bibinfo {author} {\bibfnamefont {J.}~\bibnamefont
  {Verbeke}}, \bibinfo {author} {\bibfnamefont {J.}~\bibnamefont {Randrup}},\
  and\ \bibinfo {author} {\bibfnamefont {R.}~\bibnamefont {Vogt}},\ }\href
  {https://doi.org/https://doi.org/10.1016/j.cpc.2015.02.002} {\bibfield
  {journal} {\bibinfo  {journal} {Computer Physics Communications}\ }\textbf
  {\bibinfo {volume} {191}},\ \bibinfo {pages} {178 } (\bibinfo {year}
  {2015})}\BibitemShut {NoStop}%
\bibitem [{\citenamefont {Verbeke}\ \emph {et~al.}(2018)\citenamefont
  {Verbeke}, \citenamefont {Randrup},\ and\ \citenamefont {Vogt}}]{FREYA2}%
  \BibitemOpen
  \bibfield  {author} {\bibinfo {author} {\bibfnamefont {J.}~\bibnamefont
  {Verbeke}}, \bibinfo {author} {\bibfnamefont {J.}~\bibnamefont {Randrup}},\
  and\ \bibinfo {author} {\bibfnamefont {R.}~\bibnamefont {Vogt}},\ }\href
  {https://doi.org/https://doi.org/10.1016/j.cpc.2017.09.006} {\bibfield
  {journal} {\bibinfo  {journal} {Computer Physics Communications}\ }\textbf
  {\bibinfo {volume} {222}},\ \bibinfo {pages} {263 } (\bibinfo {year}
  {2018})}\BibitemShut {NoStop}%
\bibitem [{\citenamefont {Litaize}\ \emph {et~al.}(2015)\citenamefont
  {Litaize}, \citenamefont {Serot},\ and\ \citenamefont {Berge}}]{FIFRELIN}%
  \BibitemOpen
  \bibfield  {author} {\bibinfo {author} {\bibfnamefont {O.}~\bibnamefont
  {Litaize}}, \bibinfo {author} {\bibfnamefont {O.}~\bibnamefont {Serot}},\
  and\ \bibinfo {author} {\bibfnamefont {L.}~\bibnamefont {Berge}},\ }\href
  {https://doi.org/10.1140/epja/i2015-15177-9} {\bibfield  {journal} {\bibinfo
  {journal} {The European Physical Journal A}\ }\textbf {\bibinfo {volume}
  {51}},\ \bibinfo {pages} {177} (\bibinfo {year} {2015})}\BibitemShut
  {NoStop}%
\bibitem [{\citenamefont {Okumura}\ \emph {et~al.}(2018)\citenamefont
  {Okumura}, \citenamefont {Kawano}, \citenamefont {Talou}, \citenamefont
  {Jaffke},\ and\ \citenamefont {Chiba}}]{Okumura2018}%
  \BibitemOpen
  \bibfield  {author} {\bibinfo {author} {\bibfnamefont {S.}~\bibnamefont
  {Okumura}}, \bibinfo {author} {\bibfnamefont {T.}~\bibnamefont {Kawano}},
  \bibinfo {author} {\bibfnamefont {P.}~\bibnamefont {Talou}}, \bibinfo
  {author} {\bibfnamefont {P.}~\bibnamefont {Jaffke}},\ and\ \bibinfo {author}
  {\bibfnamefont {S.}~\bibnamefont {Chiba}},\ }\href
  {https://doi.org/10.1080/00223131.2018.1467288} {\bibfield  {journal}
  {\bibinfo  {journal} {J. Nucl. Sci. Tech.}\ }\textbf {\bibinfo {volume}
  {55}},\ \bibinfo {pages} {1009} (\bibinfo {year} {2018})}\BibitemShut
  {NoStop}%
\bibitem [{\citenamefont {Lovell}\ \emph {et~al.}(2021)\citenamefont {Lovell},
  \citenamefont {Kawano}, \citenamefont {Okumura}, \citenamefont {Stetcu},
  \citenamefont {Mumpower},\ and\ \citenamefont {Talou}}]{BeoH}%
  \BibitemOpen
  \bibfield  {author} {\bibinfo {author} {\bibfnamefont {A.~E.}\ \bibnamefont
  {Lovell}}, \bibinfo {author} {\bibfnamefont {T.}~\bibnamefont {Kawano}},
  \bibinfo {author} {\bibfnamefont {S.}~\bibnamefont {Okumura}}, \bibinfo
  {author} {\bibfnamefont {I.}~\bibnamefont {Stetcu}}, \bibinfo {author}
  {\bibfnamefont {M.~R.}\ \bibnamefont {Mumpower}},\ and\ \bibinfo {author}
  {\bibfnamefont {P.}~\bibnamefont {Talou}},\ }\href
  {https://doi.org/10.1103/PhysRevC.103.014615} {\bibfield  {journal} {\bibinfo
   {journal} {Phys. Rev. C}\ }\textbf {\bibinfo {volume} {103}},\ \bibinfo
  {pages} {014615} (\bibinfo {year} {2021})}\BibitemShut {NoStop}%
\bibitem [{\citenamefont {{Geppert-Kleinrath}}\ \emph
  {et~al.}(2019)\citenamefont {{Geppert-Kleinrath}}, \citenamefont
  {{Tovesson}}, \citenamefont {{Barrett}}, \citenamefont {{Bowden}},
  \citenamefont {{Bundgaard}}, \citenamefont {{Casperson}}, \citenamefont
  {{Cebra}}, \citenamefont {{Classen}}, \citenamefont {{Cunningham}},
  \citenamefont {{Duke}}, \citenamefont {{Gearhart}}, \citenamefont {{Greife}},
  \citenamefont {{Guardincerri}}, \citenamefont {{Hagmann}}, \citenamefont
  {{Heffner}}, \citenamefont {{Hensle}}, \citenamefont {{Higgins}},
  \citenamefont {{Isenhower}}, \citenamefont {{King}}, \citenamefont {{Klay}},
  \citenamefont {{Loveland}}, \citenamefont {{Magee}}, \citenamefont
  {{Manning}}, \citenamefont {{Mendenhall}}, \citenamefont {{Ruz}},
  \citenamefont {{Sangiorgio}}, \citenamefont {{Schmitt}}, \citenamefont
  {{Seilhan}}, \citenamefont {{Snyder}}, \citenamefont {{Tate}}, \citenamefont
  {{Towell}}, \citenamefont {{Walsh}}, \citenamefont {{Watson}}, \citenamefont
  {{Yao}}, \citenamefont {{Younes}}, \citenamefont {{Leeb}},\ and\
  \citenamefont {{Niffte Collaboration}}}]{Geppert2019}%
  \BibitemOpen
  \bibfield  {author} {\bibinfo {author} {\bibfnamefont {V.}~\bibnamefont
  {{Geppert-Kleinrath}}}, \bibinfo {author} {\bibfnamefont {F.}~\bibnamefont
  {{Tovesson}}}, \bibinfo {author} {\bibfnamefont {J.~S.}\ \bibnamefont
  {{Barrett}}}, \bibinfo {author} {\bibfnamefont {N.~S.}\ \bibnamefont
  {{Bowden}}}, \bibinfo {author} {\bibfnamefont {J.}~\bibnamefont
  {{Bundgaard}}}, \bibinfo {author} {\bibfnamefont {R.~J.}\ \bibnamefont
  {{Casperson}}}, \bibinfo {author} {\bibfnamefont {D.~A.}\ \bibnamefont
  {{Cebra}}}, \bibinfo {author} {\bibfnamefont {T.}~\bibnamefont {{Classen}}},
  \bibinfo {author} {\bibfnamefont {M.}~\bibnamefont {{Cunningham}}}, \bibinfo
  {author} {\bibfnamefont {D.~L.}\ \bibnamefont {{Duke}}}, \bibinfo {author}
  {\bibfnamefont {J.}~\bibnamefont {{Gearhart}}}, \bibinfo {author}
  {\bibfnamefont {U.}~\bibnamefont {{Greife}}}, \bibinfo {author}
  {\bibfnamefont {E.}~\bibnamefont {{Guardincerri}}}, \bibinfo {author}
  {\bibfnamefont {C.}~\bibnamefont {{Hagmann}}}, \bibinfo {author}
  {\bibfnamefont {M.}~\bibnamefont {{Heffner}}}, \bibinfo {author}
  {\bibfnamefont {D.}~\bibnamefont {{Hensle}}}, \bibinfo {author}
  {\bibfnamefont {D.}~\bibnamefont {{Higgins}}}, \bibinfo {author}
  {\bibfnamefont {L.~D.}\ \bibnamefont {{Isenhower}}}, \bibinfo {author}
  {\bibfnamefont {J.}~\bibnamefont {{King}}}, \bibinfo {author} {\bibfnamefont
  {J.~L.}\ \bibnamefont {{Klay}}}, \bibinfo {author} {\bibfnamefont
  {W.}~\bibnamefont {{Loveland}}}, \bibinfo {author} {\bibfnamefont {J.~A.}\
  \bibnamefont {{Magee}}}, \bibinfo {author} {\bibfnamefont {B.}~\bibnamefont
  {{Manning}}}, \bibinfo {author} {\bibfnamefont {M.~P.}\ \bibnamefont
  {{Mendenhall}}}, \bibinfo {author} {\bibfnamefont {J.}~\bibnamefont {{Ruz}}},
  \bibinfo {author} {\bibfnamefont {S.}~\bibnamefont {{Sangiorgio}}}, \bibinfo
  {author} {\bibfnamefont {K.~T.}\ \bibnamefont {{Schmitt}}}, \bibinfo {author}
  {\bibfnamefont {B.}~\bibnamefont {{Seilhan}}}, \bibinfo {author}
  {\bibfnamefont {L.}~\bibnamefont {{Snyder}}}, \bibinfo {author}
  {\bibfnamefont {A.~C.}\ \bibnamefont {{Tate}}}, \bibinfo {author}
  {\bibfnamefont {R.~S.}\ \bibnamefont {{Towell}}}, \bibinfo {author}
  {\bibfnamefont {N.}~\bibnamefont {{Walsh}}}, \bibinfo {author} {\bibfnamefont
  {S.}~\bibnamefont {{Watson}}}, \bibinfo {author} {\bibfnamefont
  {L.}~\bibnamefont {{Yao}}}, \bibinfo {author} {\bibfnamefont
  {W.}~\bibnamefont {{Younes}}}, \bibinfo {author} {\bibfnamefont
  {H.}~\bibnamefont {{Leeb}}},\ and\ \bibinfo {author} {\bibnamefont {{Niffte
  Collaboration}}},\ }\href {https://doi.org/10.1103/PhysRevC.99.064619}
  {\bibfield  {journal} {\bibinfo  {journal} {Physics Review C}\ }\textbf
  {\bibinfo {volume} {99}},\ \bibinfo {eid} {064619} (\bibinfo {year}
  {2019})},\ \Eprint {https://arxiv.org/abs/1710.00973} {arXiv:1710.00973
  [nucl-ex]} \BibitemShut {NoStop}%
\bibitem [{\citenamefont {{G{\"o}{\"o}k}}\ \emph {et~al.}(2014)\citenamefont
  {{G{\"o}{\"o}k}}, \citenamefont {{Hambsch}},\ and\ \citenamefont
  {{Vidali}}}]{Gook2014}%
  \BibitemOpen
  \bibfield  {author} {\bibinfo {author} {\bibfnamefont {A.}~\bibnamefont
  {{G{\"o}{\"o}k}}}, \bibinfo {author} {\bibfnamefont {F.~J.}\ \bibnamefont
  {{Hambsch}}},\ and\ \bibinfo {author} {\bibfnamefont {M.}~\bibnamefont
  {{Vidali}}},\ }\href {https://doi.org/10.1103/PhysRevC.90.064611} {\bibfield
  {journal} {\bibinfo  {journal} {Physics Review C}\ }\textbf {\bibinfo
  {volume} {90}},\ \bibinfo {eid} {064611} (\bibinfo {year}
  {2014})}\BibitemShut {NoStop}%
\bibitem [{\citenamefont {{Hambsch}}\ and\ \citenamefont
  {{Oberstedt}}(1997)}]{Hambsch1997}%
  \BibitemOpen
  \bibfield  {author} {\bibinfo {author} {\bibfnamefont {F.~J.}\ \bibnamefont
  {{Hambsch}}}\ and\ \bibinfo {author} {\bibfnamefont {S.}~\bibnamefont
  {{Oberstedt}}},\ }\href {https://doi.org/10.1016/S0375-9474(97)00040-7}
  {\bibfield  {journal} {\bibinfo  {journal} {Nuclear Physics A}\ }\textbf
  {\bibinfo {volume} {617}},\ \bibinfo {pages} {347} (\bibinfo {year}
  {1997})}\BibitemShut {NoStop}%
\bibitem [{\citenamefont {{Hensle}}\ \emph {et~al.}(2020)\citenamefont
  {{Hensle}}, \citenamefont {{Barker}}, \citenamefont {{Barrett}},
  \citenamefont {{Bowden}}, \citenamefont {{Brewster}}, \citenamefont
  {{Bundgaard}}, \citenamefont {{Case}}, \citenamefont {{Casperson}},
  \citenamefont {{Cebra}}, \citenamefont {{Classen}}, \citenamefont {{Duke}},
  \citenamefont {{Fotiadis}}, \citenamefont {{Gearhart}}, \citenamefont
  {{Geppert-Kleinrath}}, \citenamefont {{Greife}}, \citenamefont
  {{Guardincerri}}, \citenamefont {{Hagmann}}, \citenamefont {{Heffner}},
  \citenamefont {{Hicks}}, \citenamefont {{Higgins}}, \citenamefont
  {{Isenhower}}, \citenamefont {{Kazkaz}}, \citenamefont {{Kemnitz}},
  \citenamefont {{Kiesling}}, \citenamefont {{King}}, \citenamefont {{Klay}},
  \citenamefont {{Latta}}, \citenamefont {{Leal}}, \citenamefont {{Loveland}},
  \citenamefont {{Lynch}}, \citenamefont {{Magee}}, \citenamefont {{Manning}},
  \citenamefont {{Mendenhall}}, \citenamefont {{Monterial}}, \citenamefont
  {{Mosby}}, \citenamefont {{Oman}}, \citenamefont {{Prokop}}, \citenamefont
  {{Sangiorgio}}, \citenamefont {{Schmitt}}, \citenamefont {{Seilhan}},
  \citenamefont {{Snyder}}, \citenamefont {{Tovesson}}, \citenamefont
  {{Towell}}, \citenamefont {{Towell}}, \citenamefont {{Towell}}, \citenamefont
  {{Walsh}}, \citenamefont {{Watson}}, \citenamefont {{Yao}}, \citenamefont
  {{Younes}},\ and\ \citenamefont {{Niffte Collaboration}}}]{Hensle2020}%
  \BibitemOpen
  \bibfield  {author} {\bibinfo {author} {\bibfnamefont {D.}~\bibnamefont
  {{Hensle}}}, \bibinfo {author} {\bibfnamefont {J.~T.}\ \bibnamefont
  {{Barker}}}, \bibinfo {author} {\bibfnamefont {J.~S.}\ \bibnamefont
  {{Barrett}}}, \bibinfo {author} {\bibfnamefont {N.~S.}\ \bibnamefont
  {{Bowden}}}, \bibinfo {author} {\bibfnamefont {K.~J.}\ \bibnamefont
  {{Brewster}}}, \bibinfo {author} {\bibfnamefont {J.}~\bibnamefont
  {{Bundgaard}}}, \bibinfo {author} {\bibfnamefont {Z.~Q.}\ \bibnamefont
  {{Case}}}, \bibinfo {author} {\bibfnamefont {R.~J.}\ \bibnamefont
  {{Casperson}}}, \bibinfo {author} {\bibfnamefont {D.~A.}\ \bibnamefont
  {{Cebra}}}, \bibinfo {author} {\bibfnamefont {T.}~\bibnamefont {{Classen}}},
  \bibinfo {author} {\bibfnamefont {D.~L.}\ \bibnamefont {{Duke}}}, \bibinfo
  {author} {\bibfnamefont {N.}~\bibnamefont {{Fotiadis}}}, \bibinfo {author}
  {\bibfnamefont {J.}~\bibnamefont {{Gearhart}}}, \bibinfo {author}
  {\bibfnamefont {V.}~\bibnamefont {{Geppert-Kleinrath}}}, \bibinfo {author}
  {\bibfnamefont {U.}~\bibnamefont {{Greife}}}, \bibinfo {author}
  {\bibfnamefont {E.}~\bibnamefont {{Guardincerri}}}, \bibinfo {author}
  {\bibfnamefont {C.}~\bibnamefont {{Hagmann}}}, \bibinfo {author}
  {\bibfnamefont {M.}~\bibnamefont {{Heffner}}}, \bibinfo {author}
  {\bibfnamefont {C.~R.}\ \bibnamefont {{Hicks}}}, \bibinfo {author}
  {\bibfnamefont {D.}~\bibnamefont {{Higgins}}}, \bibinfo {author}
  {\bibfnamefont {L.~D.}\ \bibnamefont {{Isenhower}}}, \bibinfo {author}
  {\bibfnamefont {K.}~\bibnamefont {{Kazkaz}}}, \bibinfo {author}
  {\bibfnamefont {A.}~\bibnamefont {{Kemnitz}}}, \bibinfo {author}
  {\bibfnamefont {K.~J.}\ \bibnamefont {{Kiesling}}}, \bibinfo {author}
  {\bibfnamefont {J.}~\bibnamefont {{King}}}, \bibinfo {author} {\bibfnamefont
  {J.~L.}\ \bibnamefont {{Klay}}}, \bibinfo {author} {\bibfnamefont
  {J.}~\bibnamefont {{Latta}}}, \bibinfo {author} {\bibfnamefont
  {E.}~\bibnamefont {{Leal}}}, \bibinfo {author} {\bibfnamefont
  {W.}~\bibnamefont {{Loveland}}}, \bibinfo {author} {\bibfnamefont
  {M.}~\bibnamefont {{Lynch}}}, \bibinfo {author} {\bibfnamefont {J.~A.}\
  \bibnamefont {{Magee}}}, \bibinfo {author} {\bibfnamefont {B.}~\bibnamefont
  {{Manning}}}, \bibinfo {author} {\bibfnamefont {M.~P.}\ \bibnamefont
  {{Mendenhall}}}, \bibinfo {author} {\bibfnamefont {M.}~\bibnamefont
  {{Monterial}}}, \bibinfo {author} {\bibfnamefont {S.}~\bibnamefont
  {{Mosby}}}, \bibinfo {author} {\bibfnamefont {G.}~\bibnamefont {{Oman}}},
  \bibinfo {author} {\bibfnamefont {C.}~\bibnamefont {{Prokop}}}, \bibinfo
  {author} {\bibfnamefont {S.}~\bibnamefont {{Sangiorgio}}}, \bibinfo {author}
  {\bibfnamefont {K.~T.}\ \bibnamefont {{Schmitt}}}, \bibinfo {author}
  {\bibfnamefont {B.}~\bibnamefont {{Seilhan}}}, \bibinfo {author}
  {\bibfnamefont {L.}~\bibnamefont {{Snyder}}}, \bibinfo {author}
  {\bibfnamefont {F.}~\bibnamefont {{Tovesson}}}, \bibinfo {author}
  {\bibfnamefont {C.~L.}\ \bibnamefont {{Towell}}}, \bibinfo {author}
  {\bibfnamefont {R.~S.}\ \bibnamefont {{Towell}}}, \bibinfo {author}
  {\bibfnamefont {T.~R.}\ \bibnamefont {{Towell}}}, \bibinfo {author}
  {\bibfnamefont {N.}~\bibnamefont {{Walsh}}}, \bibinfo {author} {\bibfnamefont
  {T.~S.}\ \bibnamefont {{Watson}}}, \bibinfo {author} {\bibfnamefont
  {L.}~\bibnamefont {{Yao}}}, \bibinfo {author} {\bibfnamefont
  {W.}~\bibnamefont {{Younes}}},\ and\ \bibinfo {author} {\bibnamefont {{Niffte
  Collaboration}}},\ }\href {https://doi.org/10.1103/PhysRevC.102.014605}
  {\bibfield  {journal} {\bibinfo  {journal} {Physics Review C}\ }\textbf
  {\bibinfo {volume} {102}},\ \bibinfo {eid} {014605} (\bibinfo {year}
  {2020})},\ \Eprint {https://arxiv.org/abs/2001.09381} {arXiv:2001.09381
  [nucl-ex]} \BibitemShut {NoStop}%
\bibitem [{\citenamefont {Jaffke}\ \emph {et~al.}(2018)\citenamefont {Jaffke},
  \citenamefont {M\"oller}, \citenamefont {Talou},\ and\ \citenamefont
  {Sierk}}]{Jaffke2018}%
  \BibitemOpen
  \bibfield  {author} {\bibinfo {author} {\bibfnamefont {P.}~\bibnamefont
  {Jaffke}}, \bibinfo {author} {\bibfnamefont {P.}~\bibnamefont {M\"oller}},
  \bibinfo {author} {\bibfnamefont {P.}~\bibnamefont {Talou}},\ and\ \bibinfo
  {author} {\bibfnamefont {A.~J.}\ \bibnamefont {Sierk}},\ }\href
  {https://doi.org/10.1103/PhysRevC.97.034608} {\bibfield  {journal} {\bibinfo
  {journal} {Phys. Rev. C}\ }\textbf {\bibinfo {volume} {97}},\ \bibinfo
  {pages} {034608} (\bibinfo {year} {2018})}\BibitemShut {NoStop}%
\bibitem [{\citenamefont {Randrup}\ \emph {et~al.}(2019)\citenamefont
  {Randrup}, \citenamefont {Talou},\ and\ \citenamefont {Vogt}}]{Randrup2019}%
  \BibitemOpen
  \bibfield  {author} {\bibinfo {author} {\bibfnamefont {J.}~\bibnamefont
  {Randrup}}, \bibinfo {author} {\bibfnamefont {P.}~\bibnamefont {Talou}},\
  and\ \bibinfo {author} {\bibfnamefont {R.}~\bibnamefont {Vogt}},\ }\href
  {https://doi.org/10.1103/PhysRevC.99.054619} {\bibfield  {journal} {\bibinfo
  {journal} {Phys. Rev. C}\ }\textbf {\bibinfo {volume} {99}},\ \bibinfo
  {pages} {054619} (\bibinfo {year} {2019})}\BibitemShut {NoStop}%
\bibitem [{\citenamefont {Schindler}\ and\ \citenamefont
  {Phillips}(2009)}]{Schindler2009}%
  \BibitemOpen
  \bibfield  {author} {\bibinfo {author} {\bibfnamefont {M.}~\bibnamefont
  {Schindler}}\ and\ \bibinfo {author} {\bibfnamefont {D.}~\bibnamefont
  {Phillips}},\ }\href
  {https://doi.org/https://doi.org/10.1016/j.aop.2008.09.003} {\bibfield
  {journal} {\bibinfo  {journal} {Annals of Physics}\ }\textbf {\bibinfo
  {volume} {324}},\ \bibinfo {pages} {682 } (\bibinfo {year}
  {2009})}\BibitemShut {NoStop}%
\bibitem [{\citenamefont {Dobaczewski}\ \emph {et~al.}(2014)\citenamefont
  {Dobaczewski}, \citenamefont {Nazarewicz},\ and\ \citenamefont
  {Reinhard}}]{Dobaczewski2014}%
  \BibitemOpen
  \bibfield  {author} {\bibinfo {author} {\bibfnamefont {J.}~\bibnamefont
  {Dobaczewski}}, \bibinfo {author} {\bibfnamefont {W.}~\bibnamefont
  {Nazarewicz}},\ and\ \bibinfo {author} {\bibfnamefont {P.-G.}\ \bibnamefont
  {Reinhard}},\ }\href {https://doi.org/10.1088/0954-3899/41/7/074001}
  {\bibfield  {journal} {\bibinfo  {journal} {Journal of Physics G: Nuclear and
  Particle Physics}\ }\textbf {\bibinfo {volume} {41}},\ \bibinfo {pages}
  {074001} (\bibinfo {year} {2014})}\BibitemShut {NoStop}%
\bibitem [{\citenamefont {Coello~P\'erez}\ and\ \citenamefont
  {Papenbrock}(2015)}]{Perez2015}%
  \BibitemOpen
  \bibfield  {author} {\bibinfo {author} {\bibfnamefont {E.~A.}\ \bibnamefont
  {Coello~P\'erez}}\ and\ \bibinfo {author} {\bibfnamefont {T.}~\bibnamefont
  {Papenbrock}},\ }\href {https://doi.org/10.1103/PhysRevC.92.064309}
  {\bibfield  {journal} {\bibinfo  {journal} {Phys. Rev. C}\ }\textbf {\bibinfo
  {volume} {92}},\ \bibinfo {pages} {064309} (\bibinfo {year}
  {2015})}\BibitemShut {NoStop}%
\bibitem [{\citenamefont {McDonnell}\ \emph {et~al.}(2015)\citenamefont
  {McDonnell}, \citenamefont {Schunck}, \citenamefont {Higdon}, \citenamefont
  {Sarich}, \citenamefont {Wild},\ and\ \citenamefont
  {Nazarewicz}}]{McDonnell2015}%
  \BibitemOpen
  \bibfield  {author} {\bibinfo {author} {\bibfnamefont {J.~D.}\ \bibnamefont
  {McDonnell}}, \bibinfo {author} {\bibfnamefont {N.}~\bibnamefont {Schunck}},
  \bibinfo {author} {\bibfnamefont {D.}~\bibnamefont {Higdon}}, \bibinfo
  {author} {\bibfnamefont {J.}~\bibnamefont {Sarich}}, \bibinfo {author}
  {\bibfnamefont {S.~M.}\ \bibnamefont {Wild}},\ and\ \bibinfo {author}
  {\bibfnamefont {W.}~\bibnamefont {Nazarewicz}},\ }\href
  {https://doi.org/10.1103/PhysRevLett.114.122501} {\bibfield  {journal}
  {\bibinfo  {journal} {Phys. Rev. Lett.}\ }\textbf {\bibinfo {volume} {114}},\
  \bibinfo {pages} {122501} (\bibinfo {year} {2015})}\BibitemShut {NoStop}%
\bibitem [{\citenamefont {Furnstahl}\ \emph
  {et~al.}(2015{\natexlab{a}})\citenamefont {Furnstahl}, \citenamefont {Klco},
  \citenamefont {Phillips},\ and\ \citenamefont {Wesolowski}}]{Furnstahl2015}%
  \BibitemOpen
  \bibfield  {author} {\bibinfo {author} {\bibfnamefont {R.~J.}\ \bibnamefont
  {Furnstahl}}, \bibinfo {author} {\bibfnamefont {N.}~\bibnamefont {Klco}},
  \bibinfo {author} {\bibfnamefont {D.~R.}\ \bibnamefont {Phillips}},\ and\
  \bibinfo {author} {\bibfnamefont {S.}~\bibnamefont {Wesolowski}},\ }\href
  {https://doi.org/10.1103/PhysRevC.92.024005} {\bibfield  {journal} {\bibinfo
  {journal} {Phys. Rev. C}\ }\textbf {\bibinfo {volume} {92}},\ \bibinfo
  {pages} {024005} (\bibinfo {year} {2015}{\natexlab{a}})}\BibitemShut
  {NoStop}%
\bibitem [{\citenamefont {Furnstahl}\ \emph
  {et~al.}(2015{\natexlab{b}})\citenamefont {Furnstahl}, \citenamefont {Klco},
  \citenamefont {Phillips},\ and\ \citenamefont {Wesolowski}}]{Furnstahl2015a}%
  \BibitemOpen
  \bibfield  {author} {\bibinfo {author} {\bibfnamefont {R.~J.}\ \bibnamefont
  {Furnstahl}}, \bibinfo {author} {\bibfnamefont {N.}~\bibnamefont {Klco}},
  \bibinfo {author} {\bibfnamefont {D.~R.}\ \bibnamefont {Phillips}},\ and\
  \bibinfo {author} {\bibfnamefont {S.}~\bibnamefont {Wesolowski}},\ }\href
  {https://doi.org/10.1103/PhysRevC.92.024005} {\bibfield  {journal} {\bibinfo
  {journal} {Phys. Rev. C}\ }\textbf {\bibinfo {volume} {92}},\ \bibinfo
  {pages} {024005} (\bibinfo {year} {2015}{\natexlab{b}})}\BibitemShut
  {NoStop}%
\bibitem [{\citenamefont {Wesolowski}\ \emph {et~al.}(2016)\citenamefont
  {Wesolowski}, \citenamefont {Klco}, \citenamefont {Furnstahl}, \citenamefont
  {Phillips},\ and\ \citenamefont {Thapaliya}}]{Wesolowski2016}%
  \BibitemOpen
  \bibfield  {author} {\bibinfo {author} {\bibfnamefont {S.}~\bibnamefont
  {Wesolowski}}, \bibinfo {author} {\bibfnamefont {N.}~\bibnamefont {Klco}},
  \bibinfo {author} {\bibfnamefont {R.~J.}\ \bibnamefont {Furnstahl}}, \bibinfo
  {author} {\bibfnamefont {D.~R.}\ \bibnamefont {Phillips}},\ and\ \bibinfo
  {author} {\bibfnamefont {A.}~\bibnamefont {Thapaliya}},\ }\href
  {https://doi.org/10.1088/0954-3899/43/7/074001} {\bibfield  {journal}
  {\bibinfo  {journal} {Journal of Physics G: Nuclear and Particle Physics}\
  }\textbf {\bibinfo {volume} {43}},\ \bibinfo {pages} {074001} (\bibinfo
  {year} {2016})}\BibitemShut {NoStop}%
\bibitem [{\citenamefont {Melendez}\ \emph {et~al.}(2017)\citenamefont
  {Melendez}, \citenamefont {Wesolowski},\ and\ \citenamefont
  {Furnstahl}}]{Melendez2017}%
  \BibitemOpen
  \bibfield  {author} {\bibinfo {author} {\bibfnamefont {J.~A.}\ \bibnamefont
  {Melendez}}, \bibinfo {author} {\bibfnamefont {S.}~\bibnamefont
  {Wesolowski}},\ and\ \bibinfo {author} {\bibfnamefont {R.~J.}\ \bibnamefont
  {Furnstahl}},\ }\href {https://doi.org/10.1103/PhysRevC.96.024003} {\bibfield
   {journal} {\bibinfo  {journal} {Phys. Rev. C}\ }\textbf {\bibinfo {volume}
  {96}},\ \bibinfo {pages} {024003} (\bibinfo {year} {2017})}\BibitemShut
  {NoStop}%
\bibitem [{\citenamefont {Lovell}\ and\ \citenamefont
  {Nunes}(2018)}]{Lovell2018}%
  \BibitemOpen
  \bibfield  {author} {\bibinfo {author} {\bibfnamefont {A.~E.}\ \bibnamefont
  {Lovell}}\ and\ \bibinfo {author} {\bibfnamefont {F.~M.}\ \bibnamefont
  {Nunes}},\ }\href {https://doi.org/10.1103/PhysRevC.97.064612} {\bibfield
  {journal} {\bibinfo  {journal} {Phys. Rev. C}\ }\textbf {\bibinfo {volume}
  {97}},\ \bibinfo {pages} {064612} (\bibinfo {year} {2018})}\BibitemShut
  {NoStop}%
\bibitem [{\citenamefont {Catacora-Rios}\ \emph {et~al.}(2019)\citenamefont
  {Catacora-Rios}, \citenamefont {King}, \citenamefont {Lovell},\ and\
  \citenamefont {Nunes}}]{CatacoraRios2019}%
  \BibitemOpen
  \bibfield  {author} {\bibinfo {author} {\bibfnamefont {M.}~\bibnamefont
  {Catacora-Rios}}, \bibinfo {author} {\bibfnamefont {G.~B.}\ \bibnamefont
  {King}}, \bibinfo {author} {\bibfnamefont {A.~E.}\ \bibnamefont {Lovell}},\
  and\ \bibinfo {author} {\bibfnamefont {F.~M.}\ \bibnamefont {Nunes}},\ }\href
  {https://doi.org/10.1103/PhysRevC.100.064615} {\bibfield  {journal} {\bibinfo
   {journal} {Phys. Rev. C}\ }\textbf {\bibinfo {volume} {100}},\ \bibinfo
  {pages} {064615} (\bibinfo {year} {2019})}\BibitemShut {NoStop}%
\bibitem [{\citenamefont {Phillips}\ \emph {et~al.}(2021)\citenamefont
  {Phillips}, \citenamefont {Furnstahl}, \citenamefont {Heinz}, \citenamefont
  {Maiti}, \citenamefont {Nazarewicz}, \citenamefont {Nunes}, \citenamefont
  {Plumlee}, \citenamefont {Pratola}, \citenamefont {Pratt}, \citenamefont
  {Viens},\ and\ \citenamefont {Wild}}]{Phillips2021}%
  \BibitemOpen
  \bibfield  {author} {\bibinfo {author} {\bibfnamefont {D.~R.}\ \bibnamefont
  {Phillips}}, \bibinfo {author} {\bibfnamefont {R.~J.}\ \bibnamefont
  {Furnstahl}}, \bibinfo {author} {\bibfnamefont {U.}~\bibnamefont {Heinz}},
  \bibinfo {author} {\bibfnamefont {T.}~\bibnamefont {Maiti}}, \bibinfo
  {author} {\bibfnamefont {W.}~\bibnamefont {Nazarewicz}}, \bibinfo {author}
  {\bibfnamefont {F.~M.}\ \bibnamefont {Nunes}}, \bibinfo {author}
  {\bibfnamefont {M.}~\bibnamefont {Plumlee}}, \bibinfo {author} {\bibfnamefont
  {M.~T.}\ \bibnamefont {Pratola}}, \bibinfo {author} {\bibfnamefont
  {S.}~\bibnamefont {Pratt}}, \bibinfo {author} {\bibfnamefont {F.~G.}\
  \bibnamefont {Viens}},\ and\ \bibinfo {author} {\bibfnamefont {S.~M.}\
  \bibnamefont {Wild}},\ }\href {https://doi.org/10.1088/1361-6471/abf1df}
  {\bibfield  {journal} {\bibinfo  {journal} {Journal of Physics G: Nuclear and
  Particle Physics}\ }\textbf {\bibinfo {volume} {48}},\ \bibinfo {pages}
  {072001} (\bibinfo {year} {2021})}\BibitemShut {NoStop}%
\bibitem [{\citenamefont {Semposki}\ \emph {et~al.}(2022)\citenamefont
  {Semposki}, \citenamefont {Furnstahl},\ and\ \citenamefont
  {Phillips}}]{Semposki2022}%
  \BibitemOpen
  \bibfield  {author} {\bibinfo {author} {\bibfnamefont {A.~C.}\ \bibnamefont
  {Semposki}}, \bibinfo {author} {\bibfnamefont {R.~J.}\ \bibnamefont
  {Furnstahl}},\ and\ \bibinfo {author} {\bibfnamefont {D.~R.}\ \bibnamefont
  {Phillips}},\ }\href {https://doi.org/10.1103/PhysRevC.106.044002} {\bibfield
   {journal} {\bibinfo  {journal} {Phys. Rev. C}\ }\textbf {\bibinfo {volume}
  {106}},\ \bibinfo {pages} {044002} (\bibinfo {year} {2022})}\BibitemShut
  {NoStop}%
\bibitem [{\citenamefont {Kejzlar}\ \emph {et~al.}(2023)\citenamefont
  {Kejzlar}, \citenamefont {Neufcourt},\ and\ \citenamefont
  {Nazarewicz}}]{Kejzlar2023}%
  \BibitemOpen
  \bibfield  {author} {\bibinfo {author} {\bibfnamefont {V.}~\bibnamefont
  {Kejzlar}}, \bibinfo {author} {\bibfnamefont {L.}~\bibnamefont {Neufcourt}},\
  and\ \bibinfo {author} {\bibfnamefont {W.}~\bibnamefont {Nazarewicz}},\
  }\href {https://doi.org/10.1038/s41598-023-46568-0} {\bibfield  {journal}
  {\bibinfo  {journal} {Scientific Reports}\ }\textbf {\bibinfo {volume}
  {13}},\ \bibinfo {pages} {19600} (\bibinfo {year} {2023})}\BibitemShut
  {NoStop}%
\bibitem [{\citenamefont {Sharma}\ \emph {et~al.}(2024)\citenamefont {Sharma},
  \citenamefont {Schunck},\ and\ \citenamefont {Wendt}}]{Sharma2024}%
  \BibitemOpen
  \bibfield  {author} {\bibinfo {author} {\bibfnamefont {A.}~\bibnamefont
  {Sharma}}, \bibinfo {author} {\bibfnamefont {N.}~\bibnamefont {Schunck}},\
  and\ \bibinfo {author} {\bibfnamefont {K.}~\bibnamefont {Wendt}},\ }\href
  {https://arxiv.org/abs/2411.15219} {\bibinfo {title} {Bayesian model mixing
  with multi-reference energy density functional}} (\bibinfo {year} {2024}),\
  \Eprint {https://arxiv.org/abs/2411.15219} {arXiv:2411.15219 [nucl-th]}
  \BibitemShut {NoStop}%
\bibitem [{\citenamefont {Amari}(1982)}]{Amari1982}%
  \BibitemOpen
  \bibfield  {author} {\bibinfo {author} {\bibfnamefont {S.-I.}\ \bibnamefont
  {Amari}},\ }\href {https://doi.org/10.1214/aos/1176345779} {\bibfield
  {journal} {\bibinfo  {journal} {Ann. Stat.}\ }\textbf {\bibinfo {volume}
  {10}},\ \bibinfo {pages} {357} (\bibinfo {year} {1982})}\BibitemShut
  {NoStop}%
\bibitem [{\citenamefont {Amari}(2016)}]{Amari2016}%
  \BibitemOpen
  \bibfield  {author} {\bibinfo {author} {\bibfnamefont {S.-I.}\ \bibnamefont
  {Amari}},\ }\href {https://doi.org/10.1007/978-4-431-55978-8} {\emph
  {\bibinfo {title} {{Information Geometry and Its Applications}}}},\ Applied
  Mathematical Sciences\ (\bibinfo  {publisher} {Springer Japan},\ \bibinfo
  {address} {Tokyo},\ \bibinfo {year} {2016})\BibitemShut {NoStop}%
\bibitem [{\citenamefont {Ma}\ \emph {et~al.}(1997)\citenamefont {Ma},
  \citenamefont {Ji},\ and\ \citenamefont {Farmer}}]{Ma1997}%
  \BibitemOpen
  \bibfield  {author} {\bibinfo {author} {\bibfnamefont {S.}~\bibnamefont
  {Ma}}, \bibinfo {author} {\bibfnamefont {C.}~\bibnamefont {Ji}},\ and\
  \bibinfo {author} {\bibfnamefont {J.}~\bibnamefont {Farmer}},\ }\href
  {https://doi.org/10.1016/S0893-6080(96)00049-4} {\bibfield  {journal}
  {\bibinfo  {journal} {Neural Networks}\ }\textbf {\bibinfo {volume} {10}},\
  \bibinfo {pages} {243} (\bibinfo {year} {1997})}\BibitemShut {NoStop}%
\bibitem [{\citenamefont {Amari}(1998)}]{Amari1998}%
  \BibitemOpen
  \bibfield  {author} {\bibinfo {author} {\bibfnamefont {S.-I.}\ \bibnamefont
  {Amari}},\ }\href {https://doi.org/10.1162/089976698300017746} {\bibfield
  {journal} {\bibinfo  {journal} {Neural Comput.}\ }\textbf {\bibinfo {volume}
  {10}},\ \bibinfo {pages} {251} (\bibinfo {year} {1998})}\BibitemShut
  {NoStop}%
\bibitem [{\citenamefont {Ruppeiner}(1979)}]{Ruppeiner1979}%
  \BibitemOpen
  \bibfield  {author} {\bibinfo {author} {\bibfnamefont {G.}~\bibnamefont
  {Ruppeiner}},\ }\href {https://doi.org/10.1103/PhysRevA.20.1608} {\bibfield
  {journal} {\bibinfo  {journal} {Physical Review A}\ }\textbf {\bibinfo
  {volume} {20}},\ \bibinfo {pages} {1608} (\bibinfo {year}
  {1979})}\BibitemShut {NoStop}%
\bibitem [{\citenamefont {Transtrum}\ \emph {et~al.}(2015)\citenamefont
  {Transtrum}, \citenamefont {Machta}, \citenamefont {Brown}, \citenamefont
  {Daniels}, \citenamefont {Myers},\ and\ \citenamefont
  {Sethna}}]{Transtrum2015}%
  \BibitemOpen
  \bibfield  {author} {\bibinfo {author} {\bibfnamefont {M.~K.}\ \bibnamefont
  {Transtrum}}, \bibinfo {author} {\bibfnamefont {B.~B.}\ \bibnamefont
  {Machta}}, \bibinfo {author} {\bibfnamefont {K.~S.}\ \bibnamefont {Brown}},
  \bibinfo {author} {\bibfnamefont {B.~C.}\ \bibnamefont {Daniels}}, \bibinfo
  {author} {\bibfnamefont {C.~R.}\ \bibnamefont {Myers}},\ and\ \bibinfo
  {author} {\bibfnamefont {J.~P.}\ \bibnamefont {Sethna}},\ }\href
  {https://doi.org/10.1063/1.4923066} {\bibfield  {journal} {\bibinfo
  {journal} {J. Chem. Phys.}\ }\textbf {\bibinfo {volume} {143}},\ \bibinfo
  {pages} {010901} (\bibinfo {year} {2015})}\BibitemShut {NoStop}%
\bibitem [{\citenamefont {Transtrum}(2016)}]{Transtrum2016}%
  \BibitemOpen
  \bibfield  {author} {\bibinfo {author} {\bibfnamefont {M.~K.}\ \bibnamefont
  {Transtrum}},\ }\href@noop {} {\bibfield  {journal} {\bibinfo  {journal}
  {arXiv e-prints}\ ,\ \bibinfo {pages} {arXiv:1605.08705}} (\bibinfo {year}
  {2016})}\BibitemShut {NoStop}%
\bibitem [{\citenamefont {Nik{\v{s}}i{\'{c}}}\ and\ \citenamefont
  {Vretenar}(2016)}]{Niksic2016}%
  \BibitemOpen
  \bibfield  {author} {\bibinfo {author} {\bibfnamefont {T.}~\bibnamefont
  {Nik{\v{s}}i{\'{c}}}}\ and\ \bibinfo {author} {\bibfnamefont
  {D.}~\bibnamefont {Vretenar}},\ }\href
  {https://doi.org/10.1103/PhysRevC.94.024333} {\bibfield  {journal} {\bibinfo
  {journal} {Phys. Rev. C}\ }\textbf {\bibinfo {volume} {94}},\ \bibinfo
  {pages} {024333} (\bibinfo {year} {2016})}\BibitemShut {NoStop}%
\bibitem [{\citenamefont {Tisani{\'{c}}}\ \emph {et~al.}(2020)\citenamefont
  {Tisani{\'{c}}}, \citenamefont {Smol{\v{c}}i{\'{c}}}, \citenamefont
  {Imbri{\v{s}}ak}, \citenamefont {Bondi}, \citenamefont {Zamorani},
  \citenamefont {Ceraj}, \citenamefont {Vardoulaki},\ and\ \citenamefont
  {Delhaize}}]{Tisanic2020}%
  \BibitemOpen
  \bibfield  {author} {\bibinfo {author} {\bibfnamefont {K.}~\bibnamefont
  {Tisani{\'{c}}}}, \bibinfo {author} {\bibfnamefont {V.}~\bibnamefont
  {Smol{\v{c}}i{\'{c}}}}, \bibinfo {author} {\bibfnamefont {M.}~\bibnamefont
  {Imbri{\v{s}}ak}}, \bibinfo {author} {\bibfnamefont {M.}~\bibnamefont
  {Bondi}}, \bibinfo {author} {\bibfnamefont {G.}~\bibnamefont {Zamorani}},
  \bibinfo {author} {\bibfnamefont {L.}~\bibnamefont {Ceraj}}, \bibinfo
  {author} {\bibfnamefont {E.}~\bibnamefont {Vardoulaki}},\ and\ \bibinfo
  {author} {\bibfnamefont {J.}~\bibnamefont {Delhaize}},\ }\href
  {https://doi.org/10.1051/0004-6361/201937114} {\bibfield  {journal} {\bibinfo
   {journal} {Astron. Astrophys.}\ }\textbf {\bibinfo {volume} {643}},\
  \bibinfo {pages} {A51} (\bibinfo {year} {2020})}\BibitemShut {NoStop}%
\bibitem [{\citenamefont {Nik{\v{s}}i{\'{c}}}\ \emph
  {et~al.}(2017)\citenamefont {Nik{\v{s}}i{\'{c}}}, \citenamefont
  {Imbri{\v{s}}ak},\ and\ \citenamefont {Vretenar}}]{Niksic2017}%
  \BibitemOpen
  \bibfield  {author} {\bibinfo {author} {\bibfnamefont {T.}~\bibnamefont
  {Nik{\v{s}}i{\'{c}}}}, \bibinfo {author} {\bibfnamefont {M.}~\bibnamefont
  {Imbri{\v{s}}ak}},\ and\ \bibinfo {author} {\bibfnamefont {D.}~\bibnamefont
  {Vretenar}},\ }\href {https://doi.org/10.1103/PhysRevC.95.054304} {\bibfield
  {journal} {\bibinfo  {journal} {Phys. Rev. C}\ }\textbf {\bibinfo {volume}
  {95}},\ \bibinfo {pages} {054304} (\bibinfo {year} {2017})}\BibitemShut
  {NoStop}%
\bibitem [{\citenamefont {{Imbri{\v{s}}ak}}\ and\ \citenamefont
  {{Nomura}}(2023{\natexlab{a}})}]{Imbrisak2023a}%
  \BibitemOpen
  \bibfield  {author} {\bibinfo {author} {\bibfnamefont {M.}~\bibnamefont
  {{Imbri{\v{s}}ak}}}\ and\ \bibinfo {author} {\bibfnamefont {K.}~\bibnamefont
  {{Nomura}}},\ }\href {https://doi.org/10.1103/PhysRevC.107.034304} {\bibfield
   {journal} {\bibinfo  {journal} {Physics Review C}\ }\textbf {\bibinfo
  {volume} {107}},\ \bibinfo {eid} {034304} (\bibinfo {year}
  {2023}{\natexlab{a}})},\ \Eprint {https://arxiv.org/abs/2211.13457}
  {arXiv:2211.13457 [nucl-th]} \BibitemShut {NoStop}%
\bibitem [{\citenamefont {{Imbri{\v{s}}ak}}\ and\ \citenamefont
  {{Nomura}}(2023{\natexlab{b}})}]{Imbrisak2023b}%
  \BibitemOpen
  \bibfield  {author} {\bibinfo {author} {\bibfnamefont {M.}~\bibnamefont
  {{Imbri{\v{s}}ak}}}\ and\ \bibinfo {author} {\bibfnamefont {K.}~\bibnamefont
  {{Nomura}}},\ }\href {https://doi.org/10.1103/PhysRevC.108.024321} {\bibfield
   {journal} {\bibinfo  {journal} {Physics Review C}\ }\textbf {\bibinfo
  {volume} {108}},\ \bibinfo {eid} {024321} (\bibinfo {year}
  {2023}{\natexlab{b}})},\ \Eprint {https://arxiv.org/abs/2306.08174}
  {arXiv:2306.08174 [nucl-th]} \BibitemShut {NoStop}%
\bibitem [{\citenamefont {Cram{\'e}r}(1999)}]{Cramer99}%
  \BibitemOpen
  \bibfield  {author} {\bibinfo {author} {\bibfnamefont {H.}~\bibnamefont
  {Cram{\'e}r}},\ }\href@noop {} {\emph {\bibinfo {title} {Mathematical methods
  of statistics}}},\ Vol.~\bibinfo {volume} {26}\ (\bibinfo  {publisher}
  {Princeton university press},\ \bibinfo {year} {1999})\BibitemShut {NoStop}%
\bibitem [{\citenamefont {{Nielsen}}(2013)}]{Nielsen2013}%
  \BibitemOpen
  \bibfield  {author} {\bibinfo {author} {\bibfnamefont {F.}~\bibnamefont
  {{Nielsen}}},\ }\href {https://doi.org/10.48550/arXiv.1301.3578} {\bibfield
  {journal} {\bibinfo  {journal} {arXiv e-prints}\ ,\ \bibinfo {eid}
  {arXiv:1301.3578}} (\bibinfo {year} {2013})},\ \Eprint
  {https://arxiv.org/abs/1301.3578} {arXiv:1301.3578 [cs.IT]} \BibitemShut
  {NoStop}%
\end{thebibliography}%

\end{document}